\documentclass[12pt]{article}

\usepackage{scicite}
\usepackage{color}
\usepackage{times}
\usepackage{graphicx}
\usepackage{braket}
\usepackage{amsmath}
\usepackage{amssymb}
\usepackage{dsfont}
\usepackage[T1]{fontenc}
\usepackage{scicite}

\newenvironment{sciabstract}{%
\begin{quote} \bf}
{\end{quote}}

\title{Moir\'e Cavity-Quantum Electrodynamics}

\author{Yu-Tong Wang$^{1,*}$, Qi-Hang Ye$^{2,*}$, Jun-Yong Yan$^{1,*}$, Yufei Qiao$^{1,2}$, Yu-Xin Liu$^{2}$,\\ Yong-Zheng Ye$^{1,3}$, Chen Chen$^{1}$, Xiao-Tian Cheng$^{1}$, Chen-Hui Li$^{1}$, Zi-Jian Zhang$^{1}$,\\ Cheng-Nian Huang$^{1}$, Yun Meng$^{4,5}$, Kai Zou$^{4,5}$, Wen-Kang Zhan$^{6,7}$, Chao Zhao$^{6,7}$,\\ Xiaolong Hu$^{4,5}$, Clarence Augustine T H Tee$^{8}$, Wei E. I. Sha$^{1}$, Zhixiang Huang$^{9}$,\\ Huiyun Liu$^{10}$, Chao-Yuan Jin$^{1,3,11}$, Lei Ying$^{2}$, and Feng Liu$^{1,3}$\\
\\
\normalsize{$^{1}$State Key Laboratory of Extreme Photonics and Instrumentation,}\\
\normalsize{College of Information Science and Electronic Engineering,}\\
\normalsize{Zhejiang University, Hangzhou 310027, China,}\\
\normalsize{$^{2}$School of Physics, Zhejiang Key Laboratory of Micro-nano }\\
\normalsize{Quantum Chips and Quantum Control, Zhejiang University, Hangzhou 310027, China,}\\
\normalsize{$^{3}$International Joint Innovation Center, Zhejiang University, Haining 314400, China,}\\
\normalsize{$^{4}$School of Precision Instrument and Optoelectronic Engineering,}\\
\normalsize{Tianjin University, Tianjin 300072, China,}\\
\normalsize{$^{5}$Key Laboratory of Optoelectronic Information Science and Technology, Ministry of Education,}\\
\normalsize{Tianjin 300072, China,}\\
\normalsize{$^{6}$Laboratory of Solid State Optoelectronics Information Technology,}\\
\normalsize{Institute of Semiconductors, Chinese Academy of Sciences, Beijing 100083, China,}\\
\normalsize{$^{7}$College of Materials Science and Opto-Electronic Technology,}\\
\normalsize{University of Chinese Academy of Science, Beijing 101804, China,}\\
\normalsize{$^{8}$College of Physics and Electrical Information Engineering,}\\
\normalsize{Zhejiang Normal University, Hangzhou 310058, China,}\\
\normalsize{$^{9}$Key Laboratory of Intelligent Computing and Signal Processing, Ministry of Education,}\\
\normalsize{Anhui University, Hefei 230039, China,}\\
\normalsize{$^{10}$Department of Electronic and Electrical Engineering,}\\
\normalsize{University College London, London WC1E 7JE, UK,}\\
\normalsize{$^{11}$ZJU-Hangzhou Global Scientific and Technological Innovation Center,}\\
\normalsize{Zhejiang University, Hangzhou, Zhejiang 311200, China}\\
\\
\normalsize{$^{*}$These three authors contributed equally}\\
\\
\normalsize{$^\ast$jincy@zju.edu.cn, leiying@zju.edu.cn, feng\_liu@zju.edu.cn}
}

\date{}

\begin{document}

\baselineskip24pt

\maketitle

\begin{sciabstract}
Quantum emitters are a key component in photonic quantum technologies. Enhancing single-photon emission by engineering their photonic environment is essential for improving overall efficiency in quantum information processing. However, this enhancement is often limited by the need for ultra-precise emitter placement within conventional photonic cavities. 
Inspired by fascinating physics of moir\'e pattern, we propose a novel multilayer moir\'e photonic crystal with a robust isolated flatband. Theoretical analysis reveals that, with nearly infinite photonic density of states, the moir\'e cavity simultaneously possesses a high Purcell factor and large tolerance over the emitter's position, breaking the constraints of conventional cavities.
We then experimentally demonstrate various cavity-quantum electrodynamic phenomena with a quantum dot in moir\'e cavity.
A large tuning range (up to 40-fold) of QD's radiative lifetime is achieved through strong Purcell enhancement and inhibition effects.
Our findings open the door for moir\'e flatband cavity-enhanced quantum light sources and quantum nodes for the quantum internet.
\end{sciabstract}

\section*{Introduction}\label{sec:intro}
Controlling individual single photons, i.e. the fundamental units of light described by Fock or number states~\cite{scully1997quantum}, generated from a quantum emitter~\cite{Tomm2021, Dudin2012,Kurtsiefer2000} is one of the major challenges in wide range from quantum optics~\cite{walmsley2015quantum} to quantum information technologies~\cite{couteau2023applications}. An efficient approach to manipulating single-photon emission rates and wave packets is by artificially modifying photonic environments surrounding quantum emitters, since the emission properties are dictated by these photonic modes. Given their ability to reshape the spatial and frequency distribution of electromagnetic waves, optical cavities stand out as the most powerful and versatile tool for coherent single-photon control, forming the field of cavity-quantum electrodynamics (cavity-QED)~\cite{walther2006cavity}.

Traditionally made of general mirrors, cavities confine light waves at various scales. At macro-scale Fabry-P\'erot cavities use traditional mirrors to trap light, while at the mesoscopic scale cavities in nanophotonics use defects in photonic crystals (PhCs) or distributed Bragg reflectors (DBRs) for realizing similar confinement. In the latter case, traditional mirrors are replaced with effective optical ``walls’’ like the PhCs with a frequency bandgap or DBRs, in both of which electromagnetic fields exponentially decay beyond the cavity boundary. These approaches have found success in various areas. In recent years, cavity-QED has delved into the intricate interplay between quantum emitters and fine-designed optical cavities, revealing a range of phenomena such as the Purcell effect in the weak coupling regime~\cite{Arcari2014,Lodahl2015,Somaschi2016}, strong coupling~\cite{Ohta2011,Reitzenstein2006,Yoshie2004}, and dipole-induced transparency~\cite{Javadi2015,Söllner2015,Waks2006}. 
In addition, more exotic cavities with specialized functions are proposed to enhance the photon emission such as photonic hyperbolic metamaterials~\cite{cortes2017super} and surface plasmon in metallic structures~\cite{PhysRevLett.106.020501} or to slow down the photon emission utilizing photonic structures with unique photonic dispersion relationships such as the specific PhCs with Dirac~\cite{PhysRevA.97.043831} or Weyl~\cite{PhysRevLett.123.173901,PhysRevLett.125.163602} dispersion relationships and near-zero index materials~\cite{PhysRevB.87.201101}.

In general, the Purcell effect predicts that, overall, the photon emission rate of a system is inversely proportional to the mode volume while directly proportional to the Q factor~\cite{Tang_Ni_Du_Srikrishna_Mazur_2022}. 

This implies that to sustain a high spontaneous emission rate from a quantum emitter, a small mode volume and a large Q factor are essential. While a large Q factor is often constrained by fabrication imperfections and the fundamental diffraction limit for many cavity designs~\cite{Vahala2003optical}, a small mode volume presents additional challenges related to extremely precise emitter positioning to maximize its exposure to the local field~\cite{badolato2005deterministic}.

Inspired by the fascinating physics of moir\'e superlattices in electronic and excitonic systems~\cite{cao2018correlated,mak2022semiconductor,andrei2021marvels}, its photonic counterpart~\cite{mao2021magic,PhysRevResearch.4.L032031,Wang2024experimental,oudich2024engineered,wang2020localization,Yu2023moire, Raun_Tang_Ni_Mazur_Hu_2023} offers the potential for confining photons due to its isolated flatband dispersion relation. This theoretically leads to an infinite photonic density of states at a fixed frequency, enabling simultaneous realization of an infinite Q factor and a large tolerance of emitter's location within the cavity. 

In this work, we propose utilizing the moir\'e flatband photonics to modify the Purcell effect and experimentally manipulate single photon emission from a semiconductor quantum dot (QD) within a robust quasi-1D triple-layer moir\'e cavity, eliminating the need for conventional mirrors and boundaries. Theoretical analysis shows that, due to its nearly infinite photonic density of states (DOS), both high Purcell factor and large tolerance over the emitter's location can be obtained simultaneously.
The formation of the flat photonic band and resulting light localization are confirmed by the photoluminescence (PL) spectra and mapping. A large tuning range (from $42 \pm 1$ to $1692 \pm 7$~ps) of the QD's radiative lifetime is achieved while scanning the detuning between the QD and the moir\'e cavity, with an experimentally realized Q factor of 3523. The QD and moir\'{e} PhC fabricated from III-V semiconductor is grown directly on silicon. Our work demonstrates cavity-QED with moir\'e PhC, opening the door towards moir\'e flatband cavity-enhanced quantum optical devices compatible with silicon photonic platform~\cite{Bao2023, Arrazola2021,Shang2022electrically,Wei2023monolithic}.

\section*{Results} \label{sec:results}

{\bf Quantum emitter in flatband photonics ---}
Here, we focus on quasi-1D systems as shown  in Figs.~\ref{fig:1}{\bf A}-{\bf C}, with their corresponding dispersion relations and DOS 
$P(\omega)$ are shown in Figs.~\ref{fig:1}{\bf D}-{\bf F}.
We suppose the photon volume is $V\approx AL$, where $A$ is the average cross-sectional area of quasi-1D structure and $L$ is the length of the photonic structure or the period of PhC. 
The spatial confinement of a single photon by generic ``mirrors'' is radically determined by the photonic DOS $P(\omega)$ of the photonic structure as the spontaneous emission rate of a quantum emitter $\Gamma$ is proportional to the local density of states (LDOS).
The photonic DOS and LDOS of a quasi-1D structure are respectively given by (see details in Supplementary Materials)
\begin{equation}\label{eq:dos}
    P(\omega)=\int_0^L  \rho(\omega,x) dx ,
\end{equation}
and 
\begin{equation}\label{eq:ldos}
   \rho(\omega_0,x) 
     \approx \frac{ \hbar }{4\pi A} \sum_{n}\int_{\left\{ k: \omega_k=\omega_0 \right\}} \frac{\left| \boldsymbol{\epsilon}_{n,k}(x)\right|^2}{v_g(k)}dk,
\end{equation}
where $v_g(k)$ is the group velocity and $\boldsymbol{\epsilon}_{n,k}(x)$ is the electric field density of the eigenmode $(n,k)$. Here, $k$ denotes the momentum, $n$ represents the photonic band index in a PhC, and $x$ represents the location of the quantum emitter. In general, the uniformity of LDOS over the spatial dimensions and the maximum LDOS are two important properties. 
The latter one is denoted by $\rho_\mathrm{m}$.
The former one, depicted by the uniformity $\bar{K}_\rho \equiv 3-\mathrm{Kurt} \rho(\omega_0,x)$, indicates the tolerance for quantum emitter placement in a photonic structure exhibiting a strong spontaneous emission rate. Here, $\mathrm{Kurt}(\cdot)$ represents the normalized kurtosis function. 

Based on the relationship in Eqs.~(\ref{eq:dos}) and~(\ref{eq:ldos}), we find that for a fixed LDOS at the quantum emitter's transition frequency $\omega_0$, there exists a trade-off between the spatial uniformity of the LDOS and the maximum LDOS $\rho_\mathrm{m}$ within a general quasi-1D photonic structure, as illustrated by the contour diagram in Fig.~\ref{fig:1}{\bf G}.
This agrees with the empirical conclusion that a defect PhC cavity with a small effective mode volume ($AL_\mathrm{eff}$) has a stronger enhancement of the spontaneous emission rate for a quantum emitter (or LDOS).
For instance, the defect PhC cavities in Figs.~\ref{fig:1}{\bf B},{\bf E} exhibit a finite LDOS at the resonant frequency $\omega_0$. The local field can be moderately enhanced by decreasing the number of filling holes from multiple holes (L20-L3) to one hole (H1). This enhancement is accompanied by a decrease in the effective photon volume and results in a reduced $\bar{K}_\rho$. The theoretical prediction can be confirmed by our numerical simulations, as shown in Fig.~\ref{fig:1}{\bf G}. A similar trend can be observed in Fabry-P\'erot cavities (see Fig.~\ref{fig:1}{\bf A},{\bf D}). 

Now, we investigate an idealized scenario: a PhC structure with an isolated flatband dispersion relation, as shown in Figs.~\ref{fig:1}{\bf C},{\bf F}. This configuration yields a 
divergent DOS in the frequency domain and the localization in real space~\cite{di2024dipole}, as illustrated in Figs.~\ref{fig:1}{\bf F} and {\bf C} respectively. This implies that the spontaneous emission rate can reach an exceptionally high value when the quantum emitter is placed in suitable locations, while the uniformity of LDOS can be maintained at a reasonable level, or in other words, the high LDOS and large mode volume can be achieved simultaneously. Then, we numerically confirm an optimally designed moir\'e PhC structure described in the following text. As shown in Fig.~\ref{fig:1}{\bf G}, this structure can exhibit a $\rho_\mathrm{m}$ that is nearly two orders of magnitude higher than that of the L20 cavity, while maintaining the same level of $\bar{K}_\rho$ to the L20 cavity (see details in Supplementary Materials).

Here, we emphasize that if the quantum emitter is located in a non-optimal regime, i.e. its LDOS $\rho(x_0)$ is not at the maximum LDOS position (as illustrated by the dashed circles in Figs.~\ref{fig:1}{\bf A}-{\bf C}), the behavior changes. For a FP cavity, the spontaneous emission rate is slightly modified by changing the quantum emitter's position. In the case of a defect cavity, if the quantum emitter is placed out of the defect, the emission is significantly suppressed. However, for an ideal isolated flatband, the emission rate of a quantum emitter can be large at most locations due to its infinite DOS at a fixed frequency $\omega_0$. On the other hand, in the frequency domain the flatband PhC exhibits maximum DOS, enabling it to function as a quantum emitter switch controlling both ultra-fast and ultra-slow photon emission. Although the moiré structure exhibits much higher positional tolerance compared to defect cavities, its unique characteristics result in relatively weak electric field intensity or LDOS near the AB nodes—where the inner and outer holes are most offset. Consequently, the enhancement of the radiative rate in these regions is minimal.

{\bf Moir\'{e} flatband cavity ---}
To study the single photon emission of a QD embedded in a flatband moir\'e PhC,  
firstly we design and fabricate a  quasi-1D moir\'e PhC structure. This is composed of two types of 1D PhCs depicted by two lines of blue and brown circles in Fig.~\ref{fig:2}{\bf A}. The lattice constants ($a_1$ and $a_2$) of the two 1D PhCs satisfy the condition $L = 13a_1 = 14a_2$, which is a key requirement for the formation of a moir\'{e} PhC. The separation between two 1D PhCs, referred to as the magic distance ($s$), determines the flatness and frequency of the resulting moir\'e flatband. This moir\'e PhC confines light waves along the axis of the 1D PhC. We further introduce a triple-layer moir\'e PhC design (see Fig.~\ref{fig:2}{\bf B}) by combining the two aforementioned structures. Such a multilayer moir\'e PhC is more robust to lattice constant variations, preserving a higher Q factor (see Fig. S6)~\cite{Hao2024robust}.

Additionally, to achieve in-plane 2D confinement, we expand 1D PhCs on both upper and lower sides (shaded areas in Fig.~\ref{fig:2}{\bf B}), providing light confinement in the longitudinal direction. Numerical simulations indicate a quality factor $Q=2.16\times10^4$ 
(see details in Supplementary Materials). 
Figure~\ref{fig:2}{\bf C} shows the scanning electron microscopy (SEM) image of the moir\'e PhC consisting of 5 unit cells (labeled `1' to `5') fabricated within a suspended gallium arsenide (GaAs) membrane containing indium gallium arsenide (InGaAs) QDs. 

It is worth noting that the entire device, including the QD and moir\'{e} PhC, is fabricated from III-V semiconductor grown directly on a silicon substrate using the molecular beam epitaxy technique (see details in Supplementary Materials). This heterogeneous integration approach is technically demanding due to the difficulty of growing high-quality crystals on a lattice-mismatched substrate. However, it is a key step towards large-scale integrated quantum photonic circuits based on mature silicon photonic platform~\cite{Bao2023, Arrazola2021,Shang2022electrically,Wei2023monolithic}, which are currently limited by the absence of high-performance deterministic quantum light sources due to the indirect bandgap of silicon.
Although the current QD emission wavelength lies above the silicon bandgap, it can be shifted to telecom bands via compositional tuning, strain engineering, and size control, as demonstrated in silicon-based epitaxial growth~\cite{Wan2019QDlaser, Zhou2020QDlaser}, ensuring compatibility with the silicon photonic platform.

To verify the dispersion relation of the designed moir\'e PhC, we perform the full-wave simulation, yielding a nearly flatband across the entire momentum space (see the orange curve in the left panel of Fig.~\ref{fig:2}{\bf D}). The existence of such a nearly flatband is further confirmed by the consistency of the calculated photonic DOS as shown in  Fig.~\ref{fig:2}{\bf D} (Num.) and the peak of the spatially integrated PL spectrum of a moir\'e PhC unit cell measured under high excitation laser power (see Fig.~\ref{fig:2}{\bf D} (Exp.)). All measurements in this work were conducted at $T=3.6$~K.

One of the most interesting consequences of the flatband is the localization of light. This phenomenon is demonstrated and cross-checked by the spatial field distribution and PL spectra. The calculated field distribution shows that each unit cell of the moir\'e PhC acts as a cavity. The light field is well confined within five unit cells (see Fig.~\ref{fig:2}{\bf E} (Num.)). This simulation accounting for the spatial resolution (1.5~um) of our optical measurement agrees well with the PL map measured by scanning the overlapping excitation and collection spots across the fabricated moir\'e PhC (see Fig.~\ref{fig:2}{\bf E} (Exp.)). In addition, the spectrally resolved PL signal acquired at the center of each moir\'e PhC unit cell exhibits a distinct peak with a maximum Q factor of 5026 at the energy of around 1.396~eV (see Fig.~\ref{fig:2}{\bf F}). Again, this confirms the light localization in the moir\'e PhC. The minor variations in resonant frequencies of moir\'e cavities are attributed to slight differences in lattice constants of each PhC unit cell caused by nanofabrication imperfections, while the variations in Q factors for different cells are affected by the boundary condition (See the simulation result in  Fig.~\ref{fig:2}{\bf E}) and the fabrication error(See Fig. S7){~\cite{Saadi2024moire}. The sufficient uniformity of moir\'e cavity modes demonstrates their high potential for constructing scalable arrays of identical cavity-enhanced quantum light sources.

{\bf Control of single photon  emission ---}
Following the characterization of the moir\'{e} PhC, we proceed to manipulate the spontaneous emission of a quantum emitter using the moir\'{e} flatband cavity. The first step is to identify a QD coupled with a moir\'{e} cavity. Fig.~\ref{fig:3}{\bf A} shows the magneto-PL spectra of a QD located in a moir\'{e} cavity. The position of the QD is indicated by the red trapezoid in the insert. The presence of both the QD emission (indicated by red dashed lines) and the moir\'{e} cavity mode (indicated by the red dotted line) in the same spectra confirms the spatial overlap between them. 
To clearly distinguish QD emission from the cavity mode, we employ a very low excitation laser power. Under this condition, the PL intensity does not exhibit significant enhancement at cavity resonance. This behavior is attributed to: (1) an excitation rate much lower than the Purcell-enhanced QD spontaneous emission rate; and (2) dominant QD emission into in-plane slab cavity modes at resonance, reducing collection efficiency in our top-side measurement configuration~\cite{Fujita2005}.
The single-photon nature of the QD emission is verified in a standard Hanbury Brown and Twiss (HBT) setup~\cite{BROWN_TWISS_1956}. 
Figure~\ref{fig:3}{\bf B} presents a typical result showing strong antibunching and a single-photon purity of $0.93 \pm 0.09$ without any background subtraction, which unambiguously proves that the PL signal originates from a quantum emitter.

Moreover, the coupling between the QD and moir\'{e} cavity can be proved by the polarization measurement. The polarization-dependent PL intensity is measured by rotating the half-wave plate angle in front of a linear polarizer (see details in Supplementary Materials). Typically, due to the Zeeman effect, the emission of an In(Ga)As QD subjected to a strong magnetic field in Faraday geometry splits into two branches with opposite circular polarizations~\cite{Baye2002}. In contrast, modified by the moir\'e cavity, here the photon emission from the upper branch in Fig.~\ref{fig:3}{\bf A} exhibits predominantly linear polarization. In particular, its polarization measured at a high magnetic field of $B=6$~T aligns well with that of the cavity mode, as shown in Fig.~\ref{fig:3}{\bf C}. The polarization of both the QD and moir\'{e} cavity mode are mainly along the longitudinal axis of the cavity, denoted as H in Fig.~\ref{fig:3}{\bf A} inset.
Therefore, this observation can be attributed to the QD-moir\'e cavity coupling, where the cavity mode dictates the polarization of the QD emission.

Finally, we demonstrate the control over the spontaneous emission of a quantum emitter by the moir\'{e} cavity. Figure.~\ref{fig:3}{\bf D} shows the time-resolved PL (TRPL) of the QD measured at different QD-cavity detunings using a superconducting nanowire single photon detector (SNSPD)~\cite{Hao2024}. At $B=7$~T where the QD and cavity are on resonance (see Fig.~\ref{fig:3}{\bf A}), the TRPL (yellow dots in Fig.~\ref{fig:3}{\bf D}) measured under longitudinal acoustic (LA) phonon-assisted excitation ~\cite{Quilter2015,Coste2023} yields a radiative lifetime $T_1$ as short as $42 \pm 1$~ps ($50 \pm 1$~ps) with (without) deconvolving the instrument response function (FWHM = $71 \pm 1$~ps). This lifetime corresponds to a 27 (22)-fold emission rate enhancement compared with the average lifetime $T^{'}_{1}=1121 \pm 3$~ps (green dots) for QD ensembles in GaAs bulk measured under above-barrier excitation. While at $B=0$~T with the QD far detuned from the moir\'{e} cavity mode, $T_1$ slows down to $1692 \pm 7$~ps (blue dots in Fig.~\ref{fig:3}{\bf B}) due to the Purcell inhibition ~\cite{Bayer2001, Englund2005, Hulet1985, Lodahl2004}. 

Figure~\ref{fig:3}{\bf E} summarizes the dependence of $T_1$ and the Purcell factor ($F_p = T^{'}_{1}/T_{1}$) on the QD-cavity detuning. The experimental data can be well-fitted using the model describing the Purcell effect~\cite{Liu2018} with the measured cavity linewidth (0.394~meV). As shown in Fig.~\ref{fig:3}{\bf E}, $T_1$ varies by more than one order of magnitude over a detuning range 1.427~meV, demonstrating the effective control over QD's spontaneous emission by the moir\'{e} cavity.
In addition, Purcell enhancement is also observed with another QD coupled to a separate moir\'e cavity (see details in the Supplementary Materials).

\section*{Discussion} 

In summary, we have investigated cavity-QED with a moir\'{e} flatband PhC containing a quantum emitter. The flatband formation in moir\'{e} PhC can be understood as a result of the interference of multiple optical modes~\cite{Talukdar2022, Nguyen2022,Ma2023}. Compared to conventional cavities, e.g., Fabry–P\'{e}rot cavity and PhC defect cavities~\cite{Tang_Ni_Du_Srikrishna_Mazur_2022}, one of the key advantages of moir\'{e} flatband PhC is the extremely high photonic LDOS. This enables efficient control over the QD's emission properties, including the polarization and radiative lifetime, as confirmed by the cavity-dominated polarization and a 40-fold tuning in radiative lifetime. This large tuning range is attributed to the pronounced Purcell enhancement and Purcell inhibition effects~\cite{Bayer2001, Hulet1985}.

Photonic bound states in the continuum (BIC) were also proposed with ultra-large DOS~\cite{Gao_2016,Hsu_2016,Marinica_2008}. Compared with BIC, the flatband formed in the moir\'e  PhC lies within a photonic bandgap, allowing the emission from a quantum emitter with finite linewidth to be coupled into the flatband mode~\cite{Hsu_2016,Koshelev_2018}. By contrast, in the case of BIC, the portion of quantum emitter's emission that is not strictly resonant with the BIC mode can leak into radiative continuum modes, limiting the efficiency of Purcell enhancement.

 As an outlook, combining the planar moir\'{e} PhC with various solid-state quantum emitters, including III-V QDs~\cite{Baye2002}, color centers in diamond~\cite{sipahigil2016integrated}, 2D materials\cite{He_2015} and perovskite nanocrystals~\cite{kaplan2023hong}, could enable the development of arrays of cavity-enhanced on-chip quantum light sources~\cite{michler2000quantum,Liu2018,Tomm2021,ding2023high, liu2019solid, Rota2024}, essential for large-scale quantum photonic circuits~\cite{Bao2023}. 
The high Purcell factor of moir\'e cavities not only enhances photon emission rates, but also improves photon indistinguishability by mitigating dephasing from phonons and charge noise. Additionally, integrating the moiré photonic crystal cavities with fast-light waveguides by optimizing the dispersion properties of both systems for efficient phase matching may provide an effective route for on-chip integration~\cite{Siampour2023observation}.
 Further improving the Q factor may achieve strong coupling between quantum emitters and flatband photonic structure (see numerical estimation in the Supplementary Materials), with potential applications including quantum gates~\cite{Hacker2016,Kim2013a}, nondestructive photon detection~\cite{Niemietz2021}, multiphoton graph states generation~\cite{Thomas2022a}, ultrafast single-photon optical switch~\cite{sun2018single} and quantum nodes for quantum internet~\cite{kimble2008quantum}.

\section*{Materials and Methods}

{\bf Wafer structure and sample fabrication ---}
The sample is fabricated on an InGaAs QD wafer grown by molecular beam epitaxy (MBE) on a Si substrate. The wafer structure and detailed fabrication process are shown in Figure S1. The quantum dots are embedded at the center of a 140-nm GaAs membrane, with a sacrificial layer positioned underneath. After cleaning the wafer with acetone and isopropanol, an EBL resist (ARP-6200.13) is spin-coated onto the surface. The moir\'e pattern is then defined in the resist using electron beam lithography (Raith VOYAGER EBL system). Next, inductively coupled plasma etching (OXFORD Plasmapro 100 Cobra 180) is employed to transfer the pattern into the GaAs layer. Finally,  wet etching is carried out to release the membrane, resulting in a suspended GaAs slab containing quantum dots.

{\bf Optical measurement ---}
The schematic of the set-up for optical measurements is presented 
in Supplementary Fig. S1. The sample is located in a 3.6 K closed-loop cryostat (Attocube attoDRY), equipped with a magnetic field coil capable of generating up to $B=9$~T of out-of-plane tunable magnetic field (Faraday geometry). For above-barrier excitation measurements, a 637 nm pump laser is used, produced by a continuous-wave (CW) diode laser (Thorlabs LP637). For p-shell excitation, the laser wavelength is set to 880~nm, generated by a tunable CW Ti:sapphire laser (M Squared Solstis). Emission signals are collected using a custom-built confocal microscope with a 0.85-N.A. objective lens. The collected photons are directed either to a spectrometer (Princeton Instruments HRS-750) with a 1800 grooves/mm grating for spectral analysis or to a superconducting nanowire single-photon detector for rapid single-photon detection. In TRPL measurements, the excitation pulse is provided by a tunable Ti:sapphire laser (Coherent Chameleon), emitting 150-fs pulses with an 80~MHz repetition rate. These pulses are then shaped to 8~ps using a home-made 4$f$ pulse shaper. The pulses are tuned to the QD LA-phonon excitation sideband to minimize state preparation time jitter~\cite{Quilter2015,Coste2023}. The emitted single photons are filtered by double bandpass filters before being sent to the SNSPD. For field spatial distribution measurements, the excitation and the collection spots are precisely aligned (spot size $\sim$1.5$~\mu$m). A two-dimensional raster scan with a step size of $\sim$80~nm is performed using the xy-piezo nanopositioners (attocube ANPx101) below the sample.

\section*{Figure Captions}

\textbf{Figure 1.} {\bf Schematics of photon emission in various photonic structures.} A quantum emitter marked by a green filled circle, respectively placed in ({\bf A}) a traditional Fabry-P\'erot cavity, ({\bf B}) a 1D defect PhC cavity, and ({\bf C}) a moir\'e PhC cavity. Green dashed circles stand for quantum emitters positioned at non-optimal sites. Black dashed double arrows represent the effective length of cavities with strong LDOS, while grey double arrows denote the full lengths of photonic structures. A quantum emitter is a quantum system with two energy levels: a ground state and an excited state, as illustrated in the inset of ({\bf A}). When the quantum system transitions from the excited state to the ground state, it emits a single photon with a spontaneous emission rate $\Gamma$. Left and right panels of ({\bf D}-{\bf F}) represent the dispersion relations and DOS $P(\omega)$ of photonic structures in ({\bf A}-{\bf C}), respectively. Green dots denote the effective modes in ({\bf A}-{\bf C}) and green dashed lines mark the transition frequency of the quantum emitter $\omega_0$. Dashed yellow curves in the right panel of ({\bf E}) denote the DOS inside the defect PhC cavity, distinguishing from those in the bandgap PhC regime marked by solid yellow curves. Dashed yellow lines denote the light cone. ({\bf G}) Schematically shows the uniformity of LDOS ($\bar{K}_\rho=3-\mathrm{Kurt}\rho(\omega_0,x)$) versus the maximum value of LDOS ($\rho_\mathrm{m} (\omega_0)$) for different fixed DOSs. Circles and squares represent the numerical results of the (L3, L5, L7, L10, L15, L20) and H1 defect PhC cavities, respectively, from top to bottom. Here, we use the L20 cavity as an analogy to the traditional Fabry-P\'erot cavity in (A). The red star represents the moir\'e PhC cavity. See numerical details in Supplementary Materials.

\textbf{Figure 2.} {\bf Design and characterization of moir\'{e} flatband cavity.} ({\bf A}) Two unit cells (gray dashed rectangles) of 1D moir\'e PhC composed of two 1D PhCs (brown and blue circles) with slightly different lattice constants ($a_1=209.1$~nm, $a_2=194.1$~nm). Other structural parameters: $d=133$~nm, $s=95$~nm, $L=2718$~nm. ({\bf B}) SEM image of a fabricated triple-layer moir\'e PhC unit cell formed by combining two unit cells shown in ({\bf A}). ({\bf C}) SEM image of a moir\'e PhC consisting of 5 unit cells (white dashed rectangles). ({\bf D}) Right: Comparison of calculated and experimentally measured photon density of states. The latter is obtained by spatially integrating PL spectra within a moir\'e PhC unit cell. The orange color indicates the flatband mode. ({\bf E}) Field spatial distribution of moir\'e flatband modes. Upper: Numerical calculation accounting for the spatial resolution of the subsequent optical measurement. Lower: PL map acquired by scanning the excitation and collection spots over the moir\'e PhC and recording the maximal PL intensity within $1.3939-1.3978$~eV. The FWHM of the excitation/collection spot is $\sim1.5$~$\mu$m. ({\bf F}) PL spectra of moir\'e cavity mode measured at centers of 5 moir\'e PhC unit cells marked in ({\bf C}) under high-power above-barrier excitation. All experiments in this study are performed at $T=3.6$~K. The Q factor of moir\'e cavity modes (1)-(5) are 3309, 3412, 5026, 3134 and 2602, respectively.

\textbf{Figure 3.} {\bf Manipulation of single photon emission from a QD in moir\'{e} flatband cavity.} ({\bf A}) Magnetic-field-dependent PL spectra of a QD and moir\'{e} cavity mode. The QD emission is split into two branches in an external magnetic field applied parallel to the QD growth axis (Faraday geometry). The higher-energy branch is tuned to be resonant with the moir\'{e} cavity mode at \textit{B} = 7~T. The inset depicts a moir\'{e} cavity composed of five superlattice periods, with a red dot marking the QD position. White arrows indicate the horizontal (H) and vertical (V) polarization directions. ({\bf B}) Second-order correlation measurement of single photon emission from the QD under p-shell excitation. The black curve is obtained after deconvolving the detection response function from the green fit, yielding a single-photon purity of $0.93 \pm 0.09$. The uncertainties correspond to one standard deviation from the fit. ({\bf C}) Polarization of the emission from the QD (green) and moir\'{e} cavity mode (red) characterized at \textit{B} = 6 T. The polarization of both the QD and moir\'{e} cavity mode are dominantly along the longitudinal direction denoted as H in Fig. 1{\bf A} inset. ({\bf D}) Time-resolved PL of the QD measured using an SNSPD. Gray: instrument response function (IRF) with a FWHM of $71 \pm 1$~ps. Blue (orange): single QD detuned (resonant) with moir\'{e} cavity mode under LA phonon-assisted excitation. Green: QD ensemble in bulk under above-bandgap excitation. Black curves: single exponential fit. ({\bf E}) QD–cavity detuning dependence of Purcell factor and QD lifetime. Solid lines: Lorentzian ﬁt with a fixed FWHM. Error bars represent the uncertainty extracted from exponential fitting.

\bibliography{ref}

\begin{thebibliography}{10}
\providecommand{\url}[1]{\texttt{#1}}
\expandafter\ifx\csname urlstyle\endcsname\relax
  \providecommand{\doi}[1]{doi:\discretionary{}{}{}#1}\else
  \providecommand{\doi}{doi:\discretionary{}{}{}\begingroup \urlstyle{rm}\Url}\fi

\bibitem{scully1997quantum}
M.~O. Scully, M.~S. Zubairy, \emph{Quantum optics} (Cambridge university press) (1997).

\bibitem{Tomm2021}
N.~Tomm, A.~Javadi, N.~O. Antoniadis, D.~Najer, M.~C. Löbl, A.~R. Korsch, R.~Schott, S.~R. Valentin, A.~D. Wieck, A.~Ludwig, R.~J. Warburton, A bright and fast source of coherent single photons. \emph{Nat. Nanotechnol.} \textbf{16}, 399--403 (2021).

\bibitem{Dudin2012}
Y.~O. Dudin, A.~Kuzmich, Strongly Interacting Rydberg Excitations of a Cold Atomic Gas. \emph{Science} \textbf{336}, 887–889 (2012).

\bibitem{Kurtsiefer2000}
C.~Kurtsiefer, S.~Mayer, P.~Zarda, H.~Weinfurter, Stable Solid-State Source of Single Photons. \emph{Phys. Rev. Lett.} \textbf{85}, 290--293 (2000).

\bibitem{walmsley2015quantum}
I.~Walmsley, Quantum optics: Science and technology in a new light. \emph{Science} \textbf{348}, 525--530 (2015).

\bibitem{couteau2023applications}
C.~Couteau, S.~Barz, T.~Durt, T.~Gerrits, J.~Huwer, R.~Prevedel, J.~Rarity, A.~Shields, G.~Weihs, Applications of single photons to quantum communication and computing. \emph{Nat. Rev. Phys.} \textbf{5}, 326--338 (2023).

\bibitem{walther2006cavity}
H.~Walther, B.~T. Varcoe, B.-G. Englert, T.~Becker, Cavity quantum electrodynamics. \emph{Rep. Prog. Phys.} \textbf{69}, 1325 (2006).

\bibitem{Arcari2014}
M.~Arcari, I.~Söllner, A.~Javadi, S.~Lindskov~Hansen, S.~Mahmoodian, J.~Liu, H.~Thyrrestrup, E.~Lee, J.~Song, S.~Stobbe, P.~Lodahl, Near-Unity Coupling Efficiency of a Quantum Emitter to a Photonic Crystal Waveguide. \emph{Phys. Rev. Lett.} \textbf{113}, 093603 (2014).

\bibitem{Lodahl2015}
P.~Lodahl, S.~Mahmoodian, S.~Stobbe, Interfacing single photons and single quantum dots with photonic nanostructures. \emph{Rev. Mod. Phys.} \textbf{87}, 347–400 (2015).

\bibitem{Somaschi2016}
N.~Somaschi, V.~Giesz, L.~De~Santis, J.~C. Loredo, M.~P. Almeida, G.~Hornecker, S.~L. Portalupi, T.~Grange, C.~Antón, J.~Demory, C.~Gómez, I.~Sagnes, N.~D. Lanzillotti-Kimura, A.~Lemaítre, A.~Auffeves, A.~G. White, L.~Lanco, P.~Senellart, Near-optimal single-photon sources in the solid state. \emph{Nat. Photonics} \textbf{10}, 340–345 (2016).

\bibitem{Ohta2011}
R.~Ohta, Y.~Ota, M.~Nomura, N.~Kumagai, S.~Ishida, S.~Iwamoto, Y.~Arakawa, Strong coupling between a photonic crystal nanobeam cavity and a single quantum dot. \emph{Appl. Phys. Lett.} \textbf{98}, 173104 (2011).

\bibitem{Reitzenstein2006}
S.~Reitzenstein, G.~Sęk, A.~Löffler, C.~Hofmann, S.~Kuhn, J.~P. Reithmaier, L.~V. Keldysh, V.~D. Kulakovskii, T.~L. Reinecke, A.~Forchel, Strong coupling in a single quantum dot semiconductor microcavity system, vol. 6115 (2006), p. 61151M.

\bibitem{Yoshie2004}
T.~Yoshie, A.~Scherer, J.~Hendrickson, G.~Khitrova, H.~M. Gibbs, G.~Rupper, C.~Ell, O.~B. Shchekin, D.~G. Deppe, Vacuum Rabi splitting with a single quantum dot in a photonic crystal nanocavity. \emph{Nature} \textbf{432}, 200–203 (2004).

\bibitem{Javadi2015}
A.~Javadi, I.~Söllner, M.~Arcari, S.~L. Hansen, L.~Midolo, S.~Mahmoodian, G.~Kiršanskė, T.~Pregnolato, E.~H. Lee, J.~D. Song, S.~Stobbe, P.~Lodahl, Single-photon non-linear optics with a quantum dot in a waveguide. \emph{Nat. Commun.} \textbf{6}, 8655 (2015).

\bibitem{Söllner2015}
I.~Söllner, S.~Mahmoodian, S.~L. Hansen, L.~Midolo, A.~Javadi, G.~Kiršanskė, T.~Pregnolato, H.~El-Ella, E.~H. Lee, J.~D. Song, S.~Stobbe, P.~Lodahl, Deterministic photon–emitter coupling in chiral photonic circuits. \emph{Nat. Nanotechnol.} \textbf{10}, 775–778 (2015).

\bibitem{Waks2006}
E.~Waks, J.~Vuckovic, Dipole Induced Transparency in Drop-Filter Cavity-Waveguide Systems. \emph{Phys. Rev. Lett.} \textbf{96}, 153601 (2006).

\bibitem{cortes2017super}
C.~L. Cortes, Z.~Jacob, Super-Coulombic atom--atom interactions in hyperbolic media. \emph{Nat. Commun.} \textbf{8}, 14144 (2017).

\bibitem{PhysRevLett.106.020501}
A.~Gonzalez-Tudela, D.~Martin-Cano, E.~Moreno, L.~Martin-Moreno, C.~Tejedor, F.~J. Garcia-Vidal, Entanglement of Two Qubits Mediated by One-Dimensional Plasmonic Waveguides. \emph{Phys. Rev. Lett.} \textbf{106}, 020501 (2011).

\bibitem{PhysRevA.97.043831}
A.~Gonz\'alez-Tudela, J.~I. Cirac, Exotic quantum dynamics and purely long-range coherent interactions in Dirac conelike baths. \emph{Phys. Rev. A} \textbf{97}, 043831 (2018).

\bibitem{PhysRevLett.123.173901}
L.~Ying, M.~Zhou, M.~Mattei, B.~Liu, P.~Campagnola, R.~H. Goldsmith, Z.~Yu, Extended Range of Dipole-Dipole Interactions in Periodically Structured Photonic Media. \emph{Phys. Rev. Lett.} \textbf{123}, 173901 (2019).

\bibitem{PhysRevLett.125.163602}
I.~n. Garc\'{\i}a-Elcano, A.~Gonz\'alez-Tudela, J.~Bravo-Abad, Tunable and Robust Long-Range Coherent Interactions between Quantum Emitters Mediated by Weyl Bound States. \emph{Phys. Rev. Lett.} \textbf{125}, 163602 (2020).

\bibitem{PhysRevB.87.201101}
R.~Fleury, A.~Al\`u, Enhanced superradiance in epsilon-near-zero plasmonic channels. \emph{Phys. Rev. B} \textbf{87}, 201101 (2013).

\bibitem{Tang_Ni_Du_Srikrishna_Mazur_2022}
H.~Tang, X.~Ni, F.~Du, V.~Srikrishna, E.~Mazur, On-chip light trapping in bilayer moiré photonic crystal slabs. \emph{Appl. Phys. Lett.} \textbf{121}, 231702 (2022).

\bibitem{Vahala2003optical}
K.~J. Vahala, Optical Microcavities. \emph{Nature} \textbf{424}, 839--846 (2003).

\bibitem{badolato2005deterministic}
A.~Badolato, K.~Hennessy, M.~Atature, J.~Dreiser, E.~Hu, P.~M. Petroff, A.~Imamoglu, Deterministic coupling of single quantum dots to single nanocavity modes. \emph{Science} \textbf{308}, 1158--1161 (2005).

\bibitem{cao2018correlated}
Y.~Cao, V.~Fatemi, A.~Demir, S.~Fang, S.~L. Tomarken, J.~Y. Luo, J.~D. Sanchez-Yamagishi, K.~Watanabe, T.~Taniguchi, E.~Kaxiras, \emph{et~al.}, Correlated insulator behaviour at half-filling in magic-angle graphene superlattices. \emph{Nature} \textbf{556}, 80--84 (2018).

\bibitem{mak2022semiconductor}
K.~F. Mak, J.~Shan, Semiconductor moir{\'e} materials. \emph{Nat. Nanotechnol.} \textbf{17}, 686--695 (2022).

\bibitem{andrei2021marvels}
E.~Y. Andrei, D.~K. Efetov, P.~Jarillo-Herrero, A.~H. MacDonald, K.~F. Mak, T.~Senthil, E.~Tutuc, A.~Yazdani, A.~F. Young, The marvels of moir{\'e} materials. \emph{Nat. Rev. Mater.} \textbf{6}, 201--206 (2021).

\bibitem{mao2021magic}
X.-R. Mao, Z.-K. Shao, H.-Y. Luan, S.-L. Wang, R.-M. Ma, Magic-angle lasers in nanostructured moir{\'e} superlattice. \emph{Nat. Nanotechnol.} \textbf{16}, 1099--1105 (2021).

\bibitem{PhysRevResearch.4.L032031}
D.~X. Nguyen, X.~Letartre, E.~Drouard, P.~Viktorovitch, H.~C. Nguyen, H.~S. Nguyen, Magic configurations in moir\'e superlattice of bilayer photonic crystals: Almost-perfect flatbands and unconventional localization. \emph{Phys. Rev. Res.} \textbf{4}, L032031 (2022).

\bibitem{Wang2024experimental}
X.~Wang, Z.~Liu, B.~Chen, G.~Qiu, D.~Wei, J.~Liu, Experimental Demonstration of High-Efficiency Harmonic Generation in Photonic Moiré Superlattice Microcavities. \emph{Nano Lett.} \textbf{24}, 11327--11333 (2024).

\bibitem{oudich2024engineered}
M.~Oudich, X.~Kong, T.~Zhang, C.~Qiu, Y.~Jing, Engineered moir{\'e} photonic and phononic superlattices. \emph{Nat. Mater.} \textbf{23}, 1169--1178 (2024).

\bibitem{wang2020localization}
P.~Wang, Y.~Zheng, X.~Chen, C.~Huang, Y.~V. Kartashov, L.~Torner, V.~V. Konotop, F.~Ye, Localization and delocalization of light in photonic moir{\'e} lattices. \emph{Nature} \textbf{577}, 42--46 (2020).

\bibitem{Yu2023moire}
D.~Yu, G.~Li, L.~Wang, D.~Leykam, L.~Yuan, X.~Chen, Moiré Lattice in One-Dimensional Synthetic Frequency Dimension. \emph{Phys. Rev. Lett.} \textbf{130}, 143801 (2023).

\bibitem{Raun_Tang_Ni_Mazur_Hu_2023}
A.~Raun, H.~Tang, X.~Ni, E.~Mazur, E.~L. Hu, GaN Magic Angle Laser in a Merged Moiré Photonic Crystal. \emph{ACS Photonics} \textbf{10}, 3001–3007 (2023).

\bibitem{Bao2023}
J.~Bao, Z.~Fu, T.~Pramanik, J.~Mao, Y.~Chi, Y.~Cao, C.~Zhai, Y.~Mao, T.~Dai, X.~Chen, X.~Jia, L.~Zhao, Y.~Zheng, B.~Tang, Z.~Li, J.~Luo, W.~Wang, Y.~Yang, Y.~Peng, D.~Liu, D.~Dai, Q.~He, A.~L. Muthali, L.~K. Oxenløwe, C.~Vigliar, S.~Paesani, H.~Hou, R.~Santagati, J.~W. Silverstone, A.~Laing, M.~G. Thompson, J.~L. O’Brien, Y.~Ding, Q.~Gong, J.~Wang, Very-large-scale integrated quantum graph photonics. \emph{Nat. Photonics} \textbf{17}, 573–581 (2023).

\bibitem{Arrazola2021}
J.~M. Arrazola, V.~Bergholm, K.~Brádler, T.~R. Bromley, M.~J. Collins, I.~Dhand, A.~Fumagalli, T.~Gerrits, A.~Goussev, L.~G. Helt, J.~Hundal, T.~Isacsson, R.~B. Israel, J.~Izaac, S.~Jahangiri, R.~Janik, N.~Killoran, S.~P. Kumar, J.~Lavoie, A.~E. Lita, D.~H. Mahler, M.~Menotti, B.~Morrison, S.~W. Nam, L.~Neuhaus, H.~Y. Qi, N.~Quesada, A.~Repingon, K.~K. Sabapathy, M.~Schuld, D.~Su, J.~Swinarton, A.~Száva, K.~Tan, P.~Tan, V.~D. Vaidya, Z.~Vernon, Z.~Zabaneh, Y.~Zhang, Quantum circuits with many photons on a programmable nanophotonic chip. \emph{Nature} \textbf{591}, 54–60 (2021).

\bibitem{Shang2022electrically}
C.~Shang, K.~Feng, E.~T. Hughes, A.~Clark, M.~Debnath, R.~Koscica, G.~Leake, J.~Herman, D.~Harame, P.~Ludewig, Y.~Wan, J.~E. Bowers, Electrically pumped quantum-dot lasers grown on 300 mm patterned Si photonic wafers. \emph{Light Sci. Appl.} \textbf{11}, 299 (2022).

\bibitem{Wei2023monolithic}
W.-Q. Wei, A.~He, B.~Yang, Z.-H. Wang, J.-Z. Huang, D.~Han, M.~Ming, X.~Guo, Y.~Su, J.-J. Zhang, T.~Wang, Monolithic integration of embedded III-V lasers on SOI. \emph{Light Sci. Appl.} \textbf{12}, 84 (2023).

\bibitem{di2024dipole}
E.~Di~Benedetto, A.~Gonzalez-Tudela, F.~Ciccarello, Dipole-dipole interactions mediated by a photonic flat band. \emph{arXiv preprint arXiv:2405.20382}  (2024).

\bibitem{Hao2024robust}
C.-Y. Hao, Z.~Zhan, P.~A. Pantaleón, J.-Q. He, Y.-X. Zhao, K.~Watanabe, T.~Taniguchi, F.~Guinea, L.~He, Robust flat bands in twisted trilayer graphene moiré quasicrystals. \emph{Nat. Commun.} \textbf{15}, 8437 (2024).

\bibitem{Wan2019QDlaser}
Y.~Wan, S.~Zhang, J.~C. Norman, M.~J. Kennedy, W.~He, S.~Liu, C.~Xiang, C.~Shang, J.-J. He, A.~C. Gossard, J.~E. Bowers, Tunable quantum dot lasers grown directly on silicon. \emph{Optica} \textbf{6}, 1394 (2019).

\bibitem{Zhou2020QDlaser}
T.~Zhou, M.~Tang, G.~Xiang, B.~Xiang, S.~Hark, M.~Martin, T.~Baron, S.~Pan, J.-S. Park, Z.~Liu, S.~Chen, Z.~Zhang, H.~Liu, Continuous-wave quantum dot photonic crystal lasers grown on on-axis Si (001). \emph{Nat. Commun.} \textbf{11}, 977 (2020).

\bibitem{Saadi2024moire}
C.~Saadi, H.~S. Nguyen, S.~Cueff, L.~Ferrier, X.~Letartre, S.~Callard, How many supercells are required for unconventional light confinement in moiré photonic lattices? \emph{Optica} \textbf{11}, 245 (2024).

\bibitem{Fujita2005}
M.~Fujita, S.~Takahashi, Y.~Tanaka, T.~Asano, S.~Noda, Simultaneous Inhibition and Redistribution of Spontaneous Light Emission in Photonic Crystals. \emph{Science} \textbf{308}, 1296--1298 (2005).

\bibitem{BROWN_TWISS_1956}
R.~H. Brown, R.~Q. Twiss, Correlation between Photons in Two Coherent Beams of Light. \emph{Nature} \textbf{177}, 27--29 (1956).

\bibitem{Baye2002}
M.~Bayer, G.~Ortner, O.~Stern, A.~Kuther, A.~A. Gorbunov, A.~Forchel, P.~Hawrylak, S.~Fafard, K.~Hinzer, T.~L. Reinecke, S.~N. Walck, J.~P. Reithmaier, F.~Klopf, F.~Schäfer, Fine structure of neutral and charged excitons in self-assembled In(Ga)As/(Al)GaAs quantum dots. \emph{Phys. Rev. B} \textbf{65}, 195315 (2002).

\bibitem{Hao2024}
Z.~Hao, K.~Zou, Y.~Meng, J.-Y. Yan, F.~Li, Y.~Huo, C.-Y. Jin, F.~Liu, T.~Descamps, A.~Iovan, V.~Zwiller, X.~Hu, High-performance eight-channel system with fractal superconducting nanowire single-photon detectors. \emph{Chip} \textbf{3}, 100087 (2024).

\bibitem{Quilter2015}
J.~Quilter, A.~Brash, F.~Liu, M.~Gl{\"a}ssl, A.~Barth, V.~Axt, A.~Ramsay, M.~Skolnick, A.~Fox, Phonon-Assisted Population Inversion of a Single InGaAs / GaAs Quantum Dot by Pulsed Laser Excitation. \emph{Phys. Rev. Lett.} \textbf{114}, 137401 (2015).

\bibitem{Coste2023}
N.~Coste, D.~A. Fioretto, N.~Belabas, S.~C. Wein, P.~Hilaire, R.~Frantzeskakis, M.~Gundin, B.~Goes, N.~Somaschi, M.~Morassi, A.~Lema{\^{i}}tre, I.~Sagnes, A.~Harouri, S.~E. Economou, A.~Auffeves, O.~Krebs, L.~Lanco, P.~Senellart, High-rate entanglement between a semiconductor spin and indistinguishable photons. \emph{Nat. Photonics} \textbf{17}, 582--587 (2023).

\bibitem{Bayer2001}
M.~Bayer, T.~L. Reinecke, F.~Weidner, A.~Larionov, A.~McDonald, A.~Forchel, Inhibition and Enhancement of the Spontaneous Emission of Quantum Dots in Structured Microresonators. \emph{Phys. Rev. Lett.} \textbf{86}, 3168--3171 (2001).

\bibitem{Englund2005}
D.~Englund, D.~Fattal, E.~Waks, G.~Solomon, B.~Zhang, T.~Nakaoka, Y.~Arakawa, Y.~Yamamoto, J.~Vučković, Controlling the Spontaneous Emission Rate of Single Quantum Dots in a Two-Dimensional Photonic Crystal. \emph{Phys. Rev. Lett.} \textbf{95}, 013904 (2005).

\bibitem{Hulet1985}
R.~G. Hulet, E.~S. Hilfer, D.~Kleppner, Inhibited Spontaneous Emission by a Rydberg Atom. \emph{Phys. Rev. Lett.} \textbf{55}, 2137--2140 (1985).

\bibitem{Lodahl2004}
P.~Lodahl, A.~Floris Van~Driel, I.~S. Nikolaev, A.~Irman, K.~Overgaag, D.~Vanmaekelbergh, W.~L. Vos, Controlling the dynamics of spontaneous emission from quantum dots by photonic crystals. \emph{Nature} \textbf{430}, 654--657 (2004).

\bibitem{Liu2018}
F.~Liu, A.~J. Brash, J.~O’Hara, L.~M. P.~P. Martins, C.~L. Phillips, R.~J. Coles, B.~Royall, E.~Clarke, C.~Bentham, N.~Prtljaga, I.~E. Itskevich, L.~R. Wilson, M.~S. Skolnick, A.~M. Fox, High Purcell factor generation of indistinguishable on-chip single photons. \emph{Nat. Nanotechnol.} \textbf{13}, 835–840 (2018).

\bibitem{Talukdar2022}
T.~H. Talukdar, A.~L. Hardison, J.~D. Ryckman, Moir{\'e} Effects in Silicon Photonic Nanowires. \emph{ACS Photonics} \textbf{9}, 1286--1294 (2022).

\bibitem{Nguyen2022}
D.~X. Nguyen, X.~Letartre, E.~Drouard, P.~Viktorovitch, H.~C. Nguyen, H.~S. Nguyen, Magic configurations in Moir{\'e} Superlattice of Bilayer Photonic Crystal: Almost-Perfect Flatbands and Unconventional Localization. \emph{Phys. Rev. Res.} \textbf{4}, L032031 (2022).

\bibitem{Ma2023}
R.-M. Ma, H.-Y. Luan, Z.-W. Zhao, W.-Z. Mao, S.-L. Wang, Y.-H. Ouyang, Z.-K. Shao, Twisted lattice nanocavity with theoretical quality factor exceeding 200 billion. \emph{Fundam. Res.} \textbf{3}, 537--543 (2023).

\bibitem{Gao_2016}
X.~Gao, C.~W. Hsu, B.~Zhen, X.~Lin, J.~D. Joannopoulos, M.~Soljačić, H.~Chen, Formation mechanism of guided resonances and bound states in the continuum in photonic crystal slabs. \emph{Sci. Rep.} \textbf{6}, 31908 (2016).

\bibitem{Hsu_2016}
C.~W. Hsu, B.~Zhen, A.~D. Stone, J.~D. Joannopoulos, M.~Soljačić, Bound states in the continuum. \emph{Nat. Rev. Mater.} \textbf{1}, 16048 (2016).

\bibitem{Marinica_2008}
D.~C. Marinica, A.~G. Borisov, S.~V. Shabanov, Bound States in the Continuum in Photonics. \emph{Phys. Rev. Lett.} \textbf{100}, 183902 (2008).

\bibitem{Koshelev_2018}
K.~Koshelev, S.~Lepeshov, M.~Liu, A.~Bogdanov, Y.~Kivshar, Asymmetric Metasurfaces with High-Q Resonances Governed by Bound States in the Continuum. \emph{Phys. Rev. Lett.} \textbf{121}, 193903 (2018).

\bibitem{sipahigil2016integrated}
A.~Sipahigil, R.~E. Evans, D.~D. Sukachev, M.~J. Burek, J.~Borregaard, M.~K. Bhaskar, C.~T. Nguyen, J.~L. Pacheco, H.~A. Atikian, C.~Meuwly, \emph{et~al.}, An integrated diamond nanophotonics platform for quantum-optical networks. \emph{Science} \textbf{354}, 847--850 (2016).

\bibitem{He_2015}
Y.-M. He, G.~Clark, J.~R. Schaibley, Y.~He, M.-C. Chen, Y.-J. Wei, X.~Ding, Q.~Zhang, W.~Yao, X.~Xu, C.-Y. Lu, J.-W. Pan, Single quantum emitters in monolayer semiconductors. \emph{Nat. Nanotechnol.} \textbf{10}, 497--502 (2015).

\bibitem{kaplan2023hong}
A.~E. Kaplan, C.~J. Krajewska, A.~H. Proppe, W.~Sun, T.~Sverko, D.~B. Berkinsky, H.~Utzat, M.~G. Bawendi, Hong--Ou--Mandel interference in colloidal CsPbBr3 perovskite nanocrystals. \emph{Nat. Photonics} \textbf{17}, 775--780 (2023).

\bibitem{michler2000quantum}
P.~Michler, A.~Kiraz, C.~Becher, W.~Schoenfeld, P.~Petroff, L.~Zhang, E.~Hu, A.~Imamoglu, A quantum dot single-photon turnstile device. \emph{Science} \textbf{290}, 2282--2285 (2000).

\bibitem{ding2023high}
X.~Ding, Y.-P. Guo, M.-C. Xu, R.-Z. Liu, G.-Y. Zou, J.-Y. Zhao, Z.-X. Ge, Q.-H. Zhang, H.-L. Liu, L.-J. Wang, M.-C. Chen, H.~Wang, Y.-M. He, Y.-H. Huo, C.-Y. Lu, J.-W. Pan, High-efficiency single-photon source above the loss-tolerant threshold for efficient linear optical quantum computing. \emph{arXiv} p. 2311.08347 (2023).

\bibitem{liu2019solid}
J.~Liu, R.~Su, Y.~Wei, B.~Yao, S.~F. C.~d. Silva, Y.~Yu, J.~Iles-Smith, K.~Srinivasan, A.~Rastelli, J.~Li, \emph{et~al.}, A solid-state source of strongly entangled photon pairs with high brightness and indistinguishability. \emph{Nat. Nanotechnol.} \textbf{14}, 586--593 (2019).

\bibitem{Rota2024}
M.~B. Rota, T.~M. Krieger, Q.~Buchinger, M.~Beccaceci, J.~Neuwirth, H.~Huet, N.~Horov{\'{a}}, G.~Lovicu, G.~Ronco, S.~F. {Covre da Silva}, G.~Pettinari, M.~Mocza{\l}a-Dusanowska, C.~Kohlberger, S.~Manna, S.~Stroj, J.~Freund, X.~Yuan, C.~Schneider, M.~Je{\v{z}}ek, S.~H{\"{o}}fling, F.~{Basso Basset}, T.~Huber-Loyola, A.~Rastelli, R.~Trotta, A source of entangled photons based on a cavity-enhanced and strain-tuned GaAs quantum dot. \emph{eLight} \textbf{4}, 13 (2024).

\bibitem{Siampour2023observation}
H.~Siampour, C.~O’Rourke, A.~J. Brash, M.~N. Makhonin, R.~Dost, D.~J. Hallett, E.~Clarke, P.~K. Patil, M.~S. Skolnick, A.~M. Fox, Observation of large spontaneous emission rate enhancement of quantum dots in a broken-symmetry slow-light waveguide. \emph{npj Quantum Inf.}  (2023).

\bibitem{Hacker2016}
B.~Hacker, S.~Welte, G.~Rempe, S.~Ritter, A photon–photon quantum gate based on a single atom in an optical resonator. \emph{Nature} \textbf{536}, 193--196 (2016).

\bibitem{Kim2013a}
H.~Kim, R.~Bose, T.~C. Shen, G.~S. Solomon, E.~Waks, A quantum logic gate between a solid-state quantum bit and a photon. \emph{Nat. Photonics} \textbf{7}, 373--377 (2013).

\bibitem{Niemietz2021}
D.~Niemietz, P.~Farrera, S.~Langenfeld, G.~Rempe, Nondestructive detection of photonic qubits. \emph{Nature} \textbf{591}, 570--574 (2021).

\bibitem{Thomas2022a}
P.~Thomas, L.~Ruscio, O.~Morin, G.~Rempe, Efficient generation of entangled multiphoton graph states from a single atom. \emph{Nature} \textbf{608}, 677--681 (2022).

\bibitem{sun2018single}
S.~Sun, H.~Kim, Z.~Luo, G.~S. Solomon, E.~Waks, A single-photon switch and transistor enabled by a solid-state quantum memory. \emph{Science} \textbf{361}, 57--60 (2018).

\bibitem{kimble2008quantum}
H.~J. Kimble, The quantum internet. \emph{Nature} \textbf{453}, 1023--1030 (2008).

\bibitem{hennessy2007quantum}
K.~Hennessy, A.~Badolato, M.~Winger, D.~Gerace, M.~Atat{\"u}re, S.~Gulde, S.~F{\"a}lt, E.~L. Hu, A.~Imamo{\u{g}}lu, Quantum nature of a strongly coupled single quantum dot--cavity system. \emph{Nature} \textbf{445}, 896--899 (2007).

\bibitem{kuruma2020surface}
K.~Kuruma, Y.~Ota, M.~Kakuda, S.~Iwamoto, Y.~Arakawa, Surface-passivated high-Q GaAs photonic crystal nanocavity with quantum dots. \emph{APL Photon.} \textbf{5} (2020).

\bibitem{bose2012low}
R.~Bose, D.~Sridharan, H.~Kim, G.~S. Solomon, E.~Waks, Low-Photon-Number Optical Switching with a Single Quantum Dot Coupled to a Photonic Crystal Cavity. \emph{Phys. Rev. Lett.} \textbf{108}, 227402 (2012).

\bibitem{volz2012ultrafast}
T.~Volz, A.~Reinhard, M.~Winger, A.~Badolato, K.~J. Hennessy, E.~L. Hu, A.~Imamo{\u{g}}lu, Ultrafast all-optical switching by single photons. \emph{Nat. Photonics} \textbf{6}, 605--609 (2012).

\bibitem{Berstermann2007}
T.~Berstermann, T.~Auer, H.~Kurtze, M.~Schwab, D.~R. Yakovlev, M.~Bayer, J.~Wiersig, C.~Gies, F.~Jahnke, D.~Reuter, A.~D. Wieck, Systematic study of carrier correlations in the electron-hole recombination dynamics of quantum dots. \emph{Phys. Rev. B} \textbf{76}, 165318 (2007).

\bibitem{Reithmaier2014}
G.~Reithmaier, F.~Flassig, P.~Hasch, S.~Lichtmannecker, K.~Müller, J.~Vučković, R.~Gross, M.~Kaniber, J.~J. Finley, A carrier relaxation bottleneck probed in single InGaAs quantum dots using integrated superconducting single photon detectors. \emph{Appl. Phys. Lett.} \textbf{105} (2014).

\bibitem{Yan2023}
J.-Y. Yan, C.~Chen, X.-D. Zhang, Y.-T. Wang, H.-G. Babin, A.~D. Wieck, A.~Ludwig, Y.~Meng, X.~Hu, H.~Duan, W.~Chen, W.~Fang, M.~Cygorek, X.~Lin, D.-W. Wang, C.-Y. Jin, F.~Liu, Coherent control of a high-orbital hole in a semiconductor quantum dot. \emph{Nat. Nanotechnol.} \textbf{18}, 1139--1146 (2023).

\bibitem{Zibik2009}
E.~A. Zibik, T.~Grange, B.~A. Carpenter, N.~E. Porter, R.~Ferreira, G.~Bastard, D.~Stehr, S.~Winnerl, M.~Helm, H.~Y. Liu, M.~S. Skolnick, L.~R. Wilson, Long lifetimes of quantum-dot intersublevel transitions in the terahertz range. \emph{Nat. Mater.} \textbf{8}, 803--807 (2009).

\end{thebibliography}
\bibliographystyle{sciencemag}

\section*{Acknowledgments}

We acknowledge Prof.~Alexander~Tartakovskii for fruitful discussion. We acknowledge the Micro/Nano Fabrication Center at Zhejiang University for their facility support and technical assistance. 

\section*{Funding}

F.L. acknowledges support from the National Key Research and Development Program of China (Nos. 2022YFA1204700, 2023YFB2806000) and the National Natural Science Foundation of China (Nos. U21A6006, 62075194). 
L.Y. acknowledges support from the National Natural Science Foundation of China (No. 12375021), the Zhejiang Provincial Natural Science Foundation of China
(No. LD25A050002), and the National Key Research and Development Program of China (No. 2022YFA1404203).
W.E.I.S. and Z.H. acknowledge support from the National Natural Science Foundation of China (Nos. U20A20164).
H.L. acknowledges support from the Engineering and Physical Sciences Research Council (EPSRC) of the United Kingdom (Nos. EP/P006973/1, EP/T028475/1, EP/X015300/1, EP/S030751/1, EP/V026496/1, EP/V006975/1).

\section*{Author Contributions Statement}

F.L. and L.Y. conceived the project. Q.-H.Y., C.-N.H. and Y.-X.L. designed the moir\'e cavity and performed simulations and theoretical derivations under the supervision of L.Y. and W.E.I.S. H.L., W.-K.Z., and C.Z. grew the wafer. Y.-T.W. and X.-T.C. fabricated the moir\'e cavity under the supervision of C.-H.L., C.-Y.J. J.-Y.Y., C.C., Y.Q., Z.-J.Z. and Y.-Z. Y. carried out the quantum optics experiments under the supervision of F.L. Y.M. and K.Z. helped set up the SNSPD system under the supervision of X.H. J.-Y.Y., Y.-T.W., Q.-H.Y., L.Y., and F.L. analyzed the data. X.H., C.A.T. Tee, and Z.H. provided supervision and expertise. Y.-T.W., Q.-H.Y., J.-Y.Y., L.Y., and F.L. wrote the manuscript with comments and inputs from all the authors.

\section*{Competing Interests Statement}
The authors declare no competing interests.

\section*{Data and Materials Availability}

All data supporting the findings of this study are included in the main text and supplementary materials.

\section*{Supplementary Materials}

Supplementary Text\\
Figs. S1 to S8\\

\clearpage
\begin{figure*}[t]
    \includegraphics[width=0.95\linewidth]{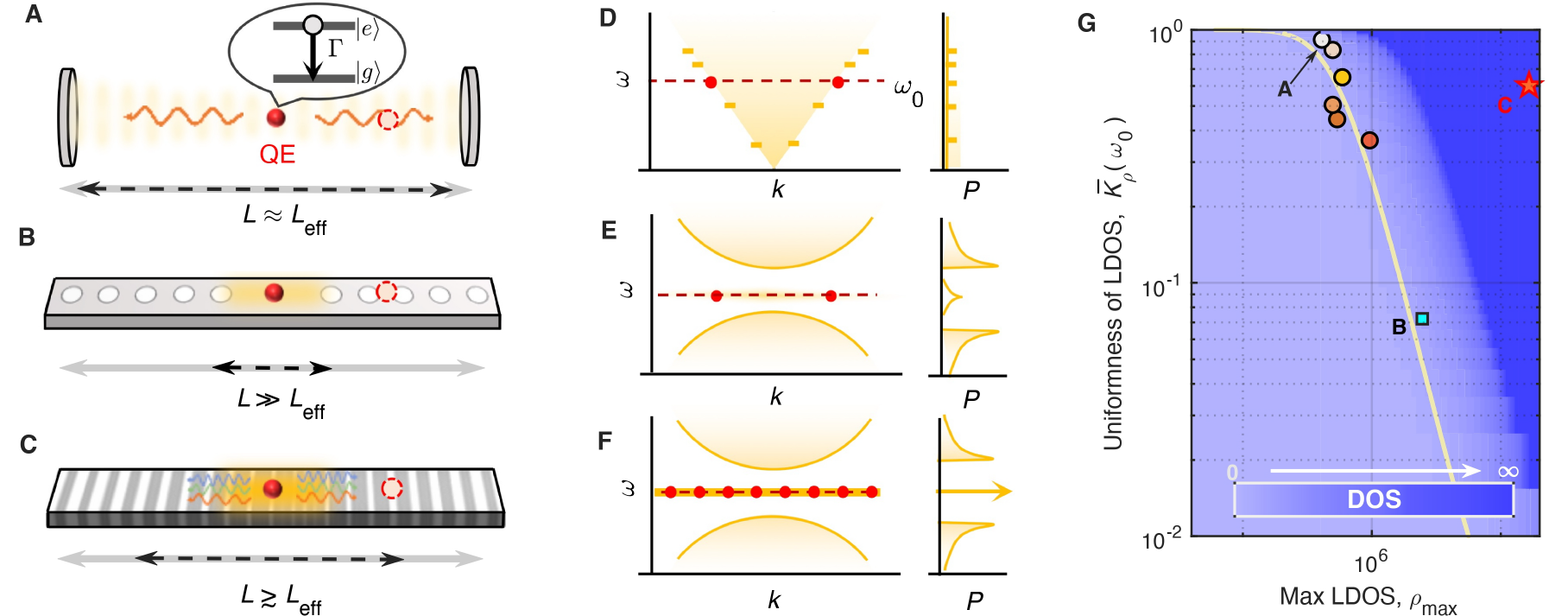}
    \caption{
    {\bf Schematics of photon emission in various photonic structures.} A quantum emitter marked by red filled circle, respectively placed in ({\bf A}) a traditional Fabry-P\'erot cavity, ({\bf B}) 1D defect PhC cavity, and  ({\bf C})   moir\'e PhC cavity. Green dashed circles stand for quantum emitters positioned at non-optimal sites. Black dashed double arrows stand for the effective length of cavities with strong LDOS, while the grey double arrows denote the full lengths of photonic structures. A quantum emitter is a quantum system with two energy levels: a ground state and an excited state, as illustrated in the inset of ({\bf A}). When the quantum system transitions from the excited state to the ground state, it emits a single photon with a spontaneous emission rate $\Gamma$.
    Left and right panels of ({\bf D}-{\bf F}) represent the dispersion relations and DOS 
    $P(\omega)$ of photonic structures in ({\bf A}-{\bf C}), respectively. Green dots denote the effective modes in  ({\bf A}-{\bf C}) and red dashed lines mark the transition frequency of the quantum emitter $\omega_0$.
    Dashed yellow curves in the right panel of ({\bf E}) denote the DOS inside the defect PhC cavity, distinguishing from those in bandgap PhC regime marked by solid yellow curves. Dashed yellow lines denote the light cone.
    ({\bf G}) Schematically shows the uniformity of LDOS ($\bar{K}_\rho=3-\mathrm{Kurt}\rho(\omega_0,x)$) versus the maximum value of LDOS ($\rho_\mathrm{m} (\omega_0)$) for different fixed DOSs. Circles and squares stand for the numerical results of the (L3, L5, L7, L10, L15, L20) and H1 defect PhC cavities, respectively, from top to bottom. Here, we use the L20 cavity as an analogy to the traditional Fabry-P\'erot cavity in (A). The red star represents the moir\'e PhC cavity. 
    See numerical details in Supplementary Materials.
    }\label{fig:1}
\end{figure*}

\begin{figure*}[ht]
    \includegraphics[width=1\linewidth]{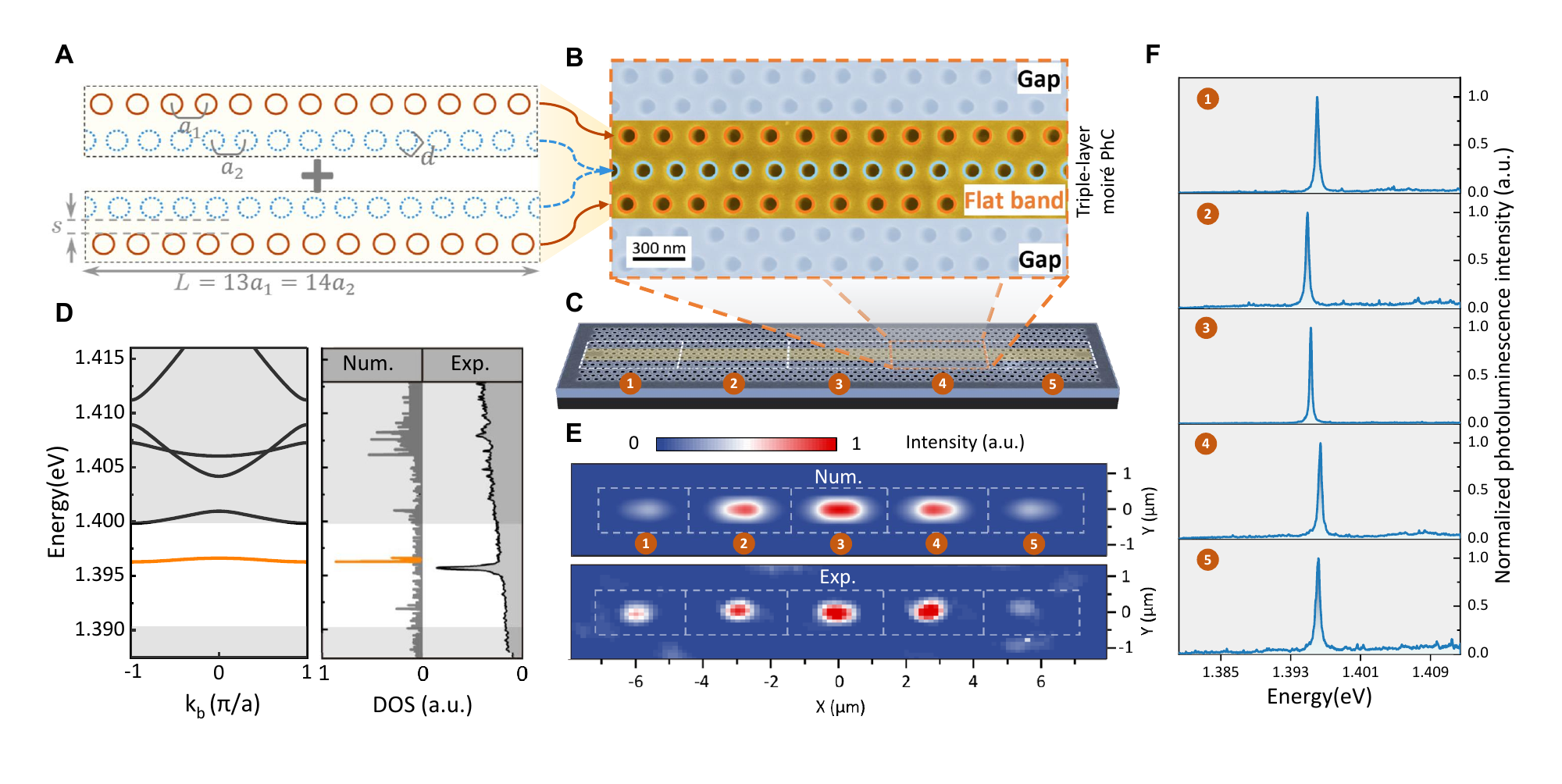}
    \caption{
    {\bf Design and characterization of moir\'{e} flatband cavity.} 
    (a) Two unit cells (gray dashed rectangles) of 1D moir\'e PhC composed of two 1D PhCs (brown and blue circles) with slightly different lattice constants ($a_1=209.1$~nm, $a_2=194.1$~nm). Other structural parameters: $d=133$~nm, $s=95$~nm, $L=2718$~nm.
    (b) SEM image of a fabricated triple-layer moir\'e PhC unit cell formed by combining two unit cells shown in (a).
    (c) SEM image of a moir\'e PhC consisting of 5 unit cells (white dashed rectangles).
    (d) Right: Comparison of calculated and experimentally measured photon density of states. The latter is obtained by spatially integrating PL spectra within a moir\'e PhC unit cell. The orange color indicates the flatband mode. 
    (e) Field spatial distribution of moir\'e flatband modes. Upper: Numerical calculation accounting for the spatial resolution of the subsequent optical measurement. Lower: PL map acquired by scanning the excitation and collection spots over the moir\'e PhC and recording the maximal PL intensity within $1.3939-1.3978$~eV. The FWHM of the excitation/collection spot is $\sim1.5$~$\mu$m.
    (f) PL spectra of moir\'e cavity mode measured at centers of 5 moir\'e PhC unit cells marked in (c) under high-power above-barrier excitation. All experiments in this study are performed at $T=3.6$~K. The Q factor of moir\'e cavity modes (1)-(5) are 3309, 3412, 5026, 3134 and 2602, respectively.}\label{fig:2}
\end{figure*}

\begin{figure*}[ht]
    \includegraphics[width=\linewidth]{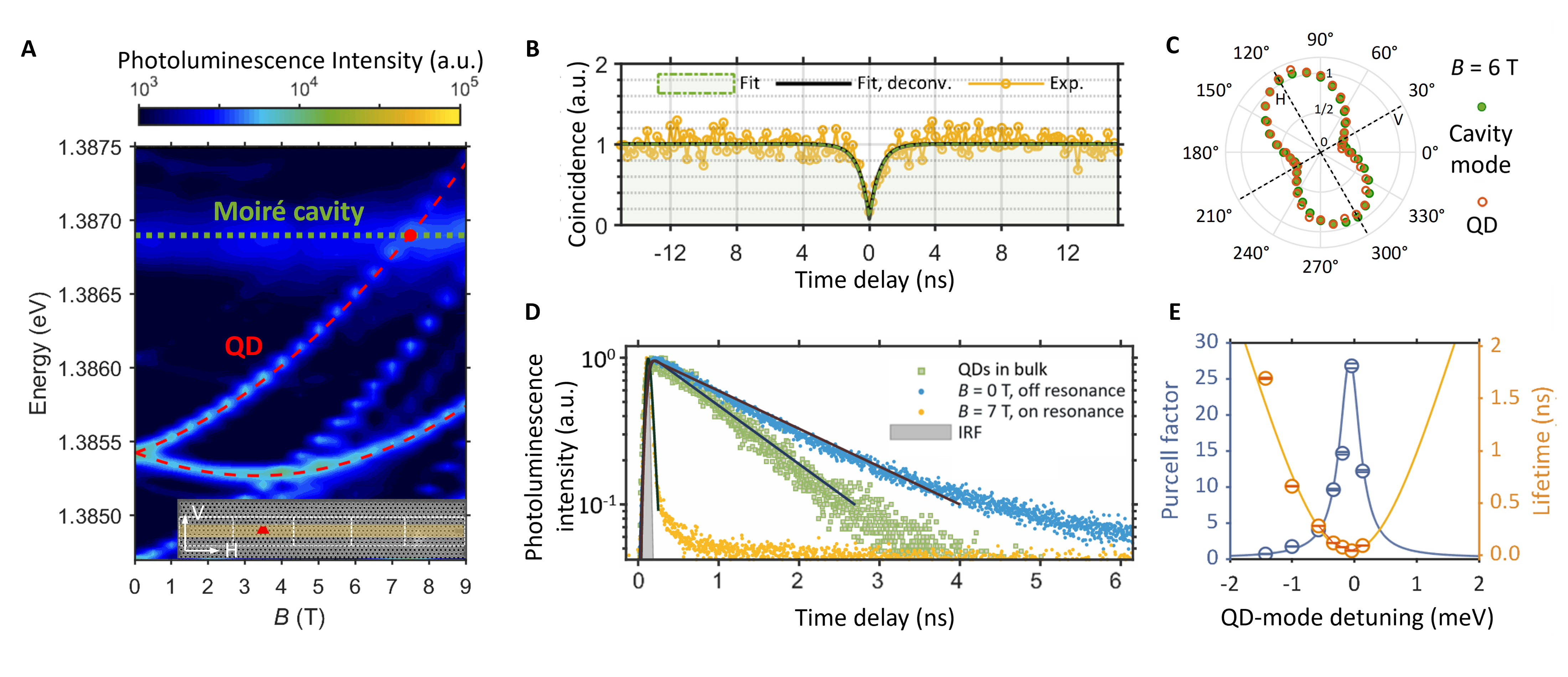}
    \caption{
    {\bf Manipulation of single photon emission from a QD in moir\'{e} flatband cavity.} 
    ({\bf A}) Magnetic-field-dependent PL spectra of a QD and moir\'{e} cavity mode. The QD emission is split into two branches in an external magnetic field applied parallel to the QD growth axis (Faraday geometry). The higher-energy branch is tuned to be resonant with the moir\'{e} cavity mode at \textit{B} = 7~T. The inset depicts a moir\'{e} cavity composed of five superlattice periods, with a red dot marking the QD position. White arrows indicate the horizontal (H) and vertical (V) polarization directions.
    ({\bf B}) Second-order correlation measurement of single photon emission from the QD under p-shell excitation. The black curve is obtained after deconvolving the detection response function from the green fit, yielding a single-photon purity of $0.93 \pm 0.09$. The uncertainties correspond to one standard deviation from the fit. 
    ({\bf C}) Polarization of the emission from the QD (green) and moir\'{e} cavity mode (red) characterized at \textit{B} = 6 T.The polarization of both the QD and moir\'{e} cavity mode are dominantly along the longitudinal direction denoted as H in Fig.~\ref{fig:3}({\bf A}) inset.
    ({\bf D}) Time-resolved PL of the QD measured using an SNSPD. Gray: instrument response function (IRF) with a FWHM of $71 \pm 1$~ps. Blue (orange): single QD detuned (resonant) with moir\'{e} cavity mode under LA phonon-assisted excitation. Green: QD ensemble in bulk under above-bandgap excitation. Black curves: single exponential fit.
    ({\bf E})  QD–cavity detuning dependence of Purcell factor and QD lifetime. Solid lines: Lorentzian ﬁt with a fixed FWHM. Error bars represent the uncertainty extracted from exponential fitting. 
    } 
    \label{fig:3}
\end{figure*}

\clearpage


\begin{center}
{
	\bf \Large 
 Supplementary Material for ``Moir\'e Cavity-Quantum Electrodynamics''
}

\vspace*{0.5cm}

Yu-Tong Wang \textit{et al.}

\vspace*{0.2cm}
{\footnotesize \em 
Corresponding author: Chao-Yuan Jin, jincy@zju.edu.cn; Lei Ying, leiying@zju.edu.cn; Feng Liu, feng\_liu@zju.edu.cn}

\end{center}

\vspace*{0.02cm}

\setcounter{equation}{0}
\setcounter{figure}{0}
\setcounter{table}{0}
\setcounter{page}{1}
\setcounter{section}{0}
\renewcommand{\theequation}{S\arabic{equation}}
\renewcommand{\thefigure}{S\arabic{figure}}
\renewcommand{\thetable}{S\arabic{table}}

\clearpage
\tableofcontents

\clearpage
\section{Experimental setup}
The experimental setup used in this work is shown in Supplementary Fig.~\ref{fig:Setup}. Measurements were conducted using a confocal microscope with the sample placed in a closed-cycle cryostat and excited by either picosecond pulses or cw lasers. The QD emission emission was collected via a single-mode fiber and directed to one of three parts: a spectrometer for the measurement of PL spectra, a TRPL setup for lifetime measurements, or an HBT setup for single-photon purity analysis. 

For polarization-dependent measurements, we excite the QD with a weak above-barrier laser to ensure a clear distinction between the cavity mode and QD emission. We rotate the HWP in front of a linear polarizer in the collection optical path (see Fig.~S1), which effectively varies the collection linear polarization basis. We then acquire a series of polarization-dependent emission spectra. By Gaussian fitting, we extract the integrated areas of the cavity and QD peaks as a function of the collection polarization angle, as shown in Fig. 3{\bf C}.

\section{Wafer structure and sample fabrication process}

Figure~\ref{fig:S2} (a) illustrates the wafer structure of the InGaAs quantum dot sample. A single layer of InGaAs QDs is at the center of a 140-nm GaAs membrane. To create a suspended membrane, a 1-$\mu$m Al\(_{0.6}\)Ga\(_{0.4}\)As sacrificial layer is grown to make the GaAs membrane suspended. Below this structure, short-period PhC (SPL) and strained layer PhC (SLS) layers facilitate the transition between the Si substrate and the III-V semiconductor.

The process for fabricating the flatband structure, depicted in Fig.~\ref{fig:S2} (b), involves the following steps: First, the pattern is defined using electron beam lithography with the photoresist ARP-6200.13, followed by development in ARP600-546 for 1 minute. Subsequently, inductively coupled plasma etching is performed with a BCl$_3$/N$_2$ ratio of 2:3 to transfer the pattern into the GaAs layer, etching to a depth of 200 nm to ensure full penetration through the membrane. The sample is then immersed in a hydrofluoric acid solution (HF:DI
=1:5) for 15 minutes to remove the sacrificial layer beneath the pattern, resulting in a suspended GaAs slab featuring a moir\'e flatband structure.

\section{Theoretical derivation}
\label{note:S2}
\subsection{Hamiltonians}

We consider a quantum emitter (QE) embedding in PhC (PhC) structure. The QE can be modeled by a two-level system and its Hamiltonian is given by
\begin{equation}
    \hat{H}_{\mathrm{QE}} = \omega_0\hat{\sigma}^{\dagger}\hat{\sigma},
\end{equation}
where $\omega_0$ is the transition frequency of quantum dot and $\hat{\sigma}^{\dagger}$ ($\hat{\sigma}$) is  the raising (lowering) operator. The Hamiltonian of PhC is written as
\begin{equation}
    \hat{H}_{\mathrm{PhC}} = \sum_{n,\mathbf{k}}\omega_{n,\mathbf{k}}\hat{a}^{\dagger}_{n,\mathbf{k}}\hat{a}_{n,\mathbf{k}},
\end{equation}
where $\omega_{n,\mathbf{k}}$ is the photon frequency for momentum $n,\mathbf{k}$. $\hat{a}^{\dagger}_{n,\mathbf{k} }$($\hat{a}_{n,\mathbf{k}}$) is the creation(annihilation) operator. The light-matter interaction term is
\begin{equation}
    \hat{H}_{\mathrm{int}} = \sum_{n,\mathbf{k}}\left[ig_{n,\mathbf{k}}(\mathbf{r})\left(\hat{\sigma}^{\dagger}+\hat{\sigma}\right)\hat{a}^{\dagger}_{k}e^{in,\mathbf{k}\cdot\mathbf{r}}+\mathrm{h.c.}\right],
\end{equation}
where $g_{n,\mathbf{k}}(r) = \sqrt{\omega_{n,\mathbf{k}}/2\epsilon_0V}\boldsymbol{\mu}\cdot\boldsymbol{\epsilon}_{n,\mathbf{k}}$ is the coupling between the photon labeled with $n,\mathbf{k}$ and the QE at position $\mathbf{r}$. Here, $\boldsymbol{\mu}$ is the dipole matrix element of the QE and $\boldsymbol{\epsilon}_\mathbf{k}$ is the electric field of the mode $\mathbf{k}$.

\subsection{Spontaneous emission rate and LDOS}
The spontaneous emission rate can be derived from perturbation theory, where the interaction is considered as the perturbation, thus the transition matrix element is given by
\begin{equation}
    M_{\mathrm{FI}} = \left\langle\mathrm{F}\right|\hat{H}_{\mathrm{int}}\left|\mathrm{I}\right\rangle + \sum_{\alpha} \frac{\left\langle\mathrm{F}\right|\hat{H}_{\mathrm{int}}\left|\mathrm{R}_{\alpha}\right\rangle\left\langle\mathrm{R}_{\alpha}\right|\hat{H}_{\mathrm{int}}\left|\mathrm{I}\right\rangle}{E_{\mathrm{I}} - E_{\mathrm{R}_\alpha}}+\cdots .
\end{equation}
For small couplings, retaining terms up to the second term of expansion already achieves very high precision. The initial state and the final state are chosen to be the same, $\left|\mathrm{I}\right\rangle = \left|\mathrm{F}\right\rangle = \left|e; 0\right\rangle$. In the bracket '$e$' means the quantum dot is at the excited state, and the Arabic number indicates the number of photons in PhC. Two intermediate states are $\left|R_1\right\rangle = \left|g; 1_{n,\mathbf{k}}\right\rangle$ and $\left|R_2\right\rangle = \left|e; 1_{n,\mathbf{k}}\right\rangle$. The energy for state $\left|\mathrm{I}\right\rangle$, $\left|R_1\right\rangle$ and $\left|R_2\right\rangle$ are respectively $E_{\mathrm{I}} = E_\mathrm{e}$, $E_{\mathrm{R}_1} = \hbar\omega_{n,\mathbf{k}}$ and $E_{\mathrm{R}_2} = E_\mathrm{e}+E_\mathrm{e}^{(n)}+\hbar\omega_{n,\mathbf{k}}$. In our discussion, The energy of $\left|g;0\right\rangle$ serves as the zero-point of energy. The final result is given by [20].
\begin{equation}
\begin{aligned}
        M_{\mathrm{FI}} = \sum_{n,\mathbf{k}  }\bigg(g_{n,\mathbf{k}}(\mathbf{r}_m) g^{*}_{n,\mathbf{k}  }(\mathbf{r}_n)\frac{1}{\omega_{n,\mathbf{k}}-\omega_0}
            +g^{*}_{n,\mathbf{k}  }(\mathbf{r}_m)g_{n,\mathbf{k}  }(\mathbf{r}_n)\frac{1}{\omega_{n,\mathbf{k}}+\omega_0}\bigg).
\end{aligned}
\end{equation} 
Replace the sum of $\mathbf{k}$ by $V/(2\pi)^3\int_{\mathrm{1BZ}} d^3\mathbf{k}$ and take the imaginary part, we obtain the spontaneous emission rate as
\begin{equation}
    \begin{aligned}
    \Gamma(\omega_0) = \sum_{n}\int_{\mathrm{1BZ}} d^3\mathbf{k}\frac{\omega_{n,\mathbf{k}}}{16\pi^2\epsilon_0}\left|\boldsymbol{\mu}\cdot\boldsymbol{\epsilon}_{\mathbf{k}}\right|^2
    \delta(\omega_{n,\mathbf{k}} - \omega_0) 
    = \sum_{n}\frac{ \omega_0}{16\pi^2\epsilon_0}\int_{\left\{ \mathbf{k}: \omega_{n,\mathbf{k}} = \omega_0 \right\} } \frac{\left|\boldsymbol{\mu}\cdot\boldsymbol{\epsilon}_{\mathbf{k}}\right|^2}{|v_g(\mathbf{k})|}dS_{\mathbf{k}}.
    \end{aligned}
\end{equation}

For the case of quasi-1D PhC structure, only $k_x$ direction has continuous dispersion relation. Thus, the integral over the iso-frequency surface is reduced to the integral over the momentum direction $k$ along the $x$ direction. Here, we use $k$ to represent the momentum along the $x$ direction. We assume that the electric field distribution at the $y-z$ cross section is uniform for each mode $k$. The cross-section area of the quasi-1D photonic structure is $A$. Then, the spontaneous emission can be re-written as
\begin{equation}
    \Gamma(\omega_0) 
     \approx \sum_{n}\frac{ \omega_0}{4A\epsilon_0}\int_{\left\{ k: \omega_k=\omega_0 \right\}} \frac{\left|\boldsymbol{\mu}\cdot\boldsymbol{\epsilon}_{n,k}\right|^2}{v_g(k)}dk.
\end{equation}

In general, the relationship between the spontaneous emission rate and the photonic local density of state is given by
\begin{equation}
   \Gamma(\boldsymbol{\mu}\rightarrow 1,\omega_0) = \sum_{n}\frac{\pi\omega_0}{\hbar\epsilon_0} \rho(\omega,\mathbf{r}), 
\end{equation}
the LDOS is given by
\begin{equation}
    \rho(\omega,\mathbf{r}) = \sum_{n}\int_{\omega_k = \omega_0}\frac{\hbar}{16\pi^2|v_g(\mathbf{k})|}|\boldsymbol{\epsilon}_{n,k}(\mathbf{r})|^2dS_{\mathbf{k}} .
\end{equation}
For quasi-1D scenario, we have 
\begin{equation}
    \rho(\omega_0,x) 
     \approx \frac{ \hbar }{4\pi A} \sum_{n}\int_{k \in \left\{ \omega_k=\omega_0 \right\}} \frac{\left| \boldsymbol{\epsilon}_{n,k}(x)\right|^2}{v_g(k)}dk.
\end{equation}

\subsection{Purcell factor}

We use the general definition to derive the Purcell factor. At first, we consider the emission power
\begin{equation}
    W = \frac{\omega}{2}\mathrm{Im}\left[\boldsymbol{\mu}\cdot \mathbf{E}(\mathbf{r}_\mathrm{s})\right],
\end{equation}
where $\boldsymbol{\mu}$ is the dipole element and $\mathbf{r}_s$ is its position. The electric field is given by the Helmholtz equation:
\begin{equation}
    \nabla\times\nabla\times\mathbf{E}(\mathbf{r})-\epsilon(\mathbf{r})k_0^2\mathbf{E}(\mathbf{r}) = \mathrm{i}\mu_0\omega\mathbf{j}(\mathbf{r}).
\end{equation}
Alternatively, we can express the electric field with the Green's function
\begin{equation}
\mathbf{E}(\mathbf{r})=\mathrm{i}\mu_0\omega\int \mathbf{G}(\mathbf{r},\mathbf{r}', \omega)\mathbf{j}(\mathbf{r}')\mathrm{d}\mathbf{r}',
\end{equation}
where the Green's function can be obtained from
\begin{equation}
    \nabla\times\nabla\times\mathbf{G}(\mathbf{r},\mathbf{r}', \omega)-\epsilon(\mathbf{r})k_0^2\mathbf{G}(\mathbf{r},\mathbf{r}', \omega)=\mathbf{I}_{3\times3}\delta(\mathbf{r}-\mathbf{r}').
\end{equation}
For a point-like quantum dipole, we have
\begin{equation}
    W = \frac{\mu_0\omega^3}{2}|\mathbf{\mu}|^2\mathrm{Im}\left[\hat{\mathbf{\mu}}\cdot\mathbf{G}(\mathbf{r}_s,\mathbf{r}_s', \omega)\cdot\hat{\mathbf{\mu}}\right).
\end{equation}
In free space, the emission power is given by
\begin{equation}
    W_0 = \frac{\omega^4}{12\pi\epsilon_0c^3}|\mathbf{\mu}|^2.
\end{equation}
Then, the Purcell factor is written as
\begin{equation}\label{eq:purcell}
    F_P = \frac{W}{W_0} = \frac{6\pi}{k_0}\mathrm{Im}\left[\hat{\mathbf{\mu}}\cdot\mathbf{G}(\mathbf{r}_s,\mathbf{r}_s', \omega)\cdot\hat{\mathbf{\mu}}\right).
\end{equation}
With the eigenmodes $\mathbf{e}_n(\mathbf{r})$ of the Helmholtz equation, we can express the Green's function as
\begin{equation}
    \mathbf{G}(\mathbf{r},\mathbf{r}', \omega)=c^2\sum_n \frac{\boldsymbol{\epsilon}_n(\mathbf{r})\otimes\boldsymbol{\epsilon}^{*}_n(\mathbf{r}')}{\omega_n^2 - \omega^2 - i\omega\gamma_n}.
\end{equation}
Here, the notation $\bigotimes$ denotes the dyadic product. $\gamma_n$ is the damping rate of mode $n$. For a periodic structure, to make the mode more explicit, we rewrite the Green's function as
\begin{equation}\label{seq:gf2}
    \mathbf{G}(\mathbf{r},\mathbf{r}', \omega)=\frac{Vc^2}{(2\pi)^3}\int  \frac{\boldsymbol{\epsilon}_{n, \mathbf{k}}(\mathbf{r})\otimes\boldsymbol{\epsilon}^{*}_{n, \mathbf{k}}(\mathbf{r}')}{\omega_{n, \mathbf{k}}^2 - \omega^2 - i\omega\gamma_{n, \mathbf{k}}}d^3\mathbf{k},
\end{equation}
where we use $n$ to labelel the eigenenergy and $\mathbf{k}$ to label the Bloch vector in the 1st Brillouin zone. As $\omega$ is close to a flatband, the summation in Eq.~(\ref{seq:gf2}) becomes extremely large and thus it will lead to strong Purcell enhancement. 

\subsection{Coupling strength}
The coupling strength between the flatband photonic mode and a quantum emitter can be calculated by
\begin{equation}
    g(\mathbf{r}) = \sum_{\omega_{\mathbf{k}}\omega_0}\frac{\boldsymbol{\mu}\cdot \boldsymbol{\epsilon}_{\mathbf{k}}(\mathbf{r})}{\hbar}
\end{equation}
As the flatband mode for different wavevectors at the same frequency exhibits a similar $\mathbf{E}$-field distribution, we can approximate the coupling strength as  
\begin{equation}
    g(\mathbf{r}) \approx N_{\mathbf{k}} \frac{\boldsymbol{\mu} \cdot \boldsymbol{\epsilon}_{\mathbf{k}}(\mathbf{r})}{\hbar},
\end{equation}
where $N_{\mathbf{k}}$ is the number of $k$ points in the first Brillouin zone.  
For a finite-size structure in the experiment, this value is proportional to the least common multiple of the two lattice periods in the moir\'e structure and the total number of unit cells.
For example, with a $31:32$ hole ratio in two columns and three unit cells, our numerical simulations predict a coupling strength $g$ of approximately $24$~GHz. This value is approaching the values achieved in the former works of L3 PhC cavities for strong coupling \cite{hennessy2007quantum,kuruma2020surface,bose2012low,volz2012ultrafast}.

\section{Numerical results for different PhC cavities}
In Fig.~\ref{fig:S3}, we present the distributions of the Purcell factor for the moir\'e structure, h1 cavity, and various L-type cavities. As shown, the moir\'e structure theoretically exhibits a significantly larger Purcell factor compared to conventional cavities, while still maintaining a considerable spatial extent. 
Theoretically, this feature breaks the conventional trade-off observed in traditional cavities, where the h1 and L3 cavities achieve relatively high Purcell factors but with small effective mode volume. Conversely, larger L-type cavities, such as L10, L15, and L20, offer a broader spatial extent while they have smaller Purcell factors. Same conclusions in LDOS distributions are shown in Fig.~\ref{fig:S4}.

Figure~\ref{fig:S6}(a) shows the bandgap area, which typically correlates with a high Q factor for the flatband within the bandgap. Figure~\ref{fig:S6}(b) presents the numerical results for the Q factor of moir\'e cavity and L3 cavity. Considering material dissipation, we find that the Q factor of the moiré cavity varies very slowly with the hole diameter, similar to the behavior observed in the defect cavity. This suggests that fabricating the moiré structure may pose additional challenges compared to the defect cavity.

Figure~\ref{fig:S7} presents the robustness analysis for the traditiona moir\'e structure and the structure designed by us. Our design consists of three lines of holes with separations $a_1$, $a_2$, and $a_1$, respectively, while the traditional two-line design consists of two lines of holes with separations $a_1$ and $a_2$. For a more practical simulation, we introduce an imaginary component to the refractive index of GaAs, which causes the Q factor to decrease by approximately three orders of magnitude. As the difference in hole separations $\Delta a = a_1 - a_2$ changes by a small amount (±2.5\%), our design shows a smaller and more linear change in the Q factor, while the traditional design exhibits a larger and non-linear change, indicating that the three-line design has lower sensitivity to the error of the $\Delta a$. We further evaluate the robustness of the structures against disorder in hole diameter $d$. As the errors in $d$ increase, the $Q$ factors of both structures decline. Notably, the $Q$ factor of our structure decreases faster than that of the traditional structure, primarily due to the larger number of holes in our design, making it more susceptible to disorder in $d$. These results suggest that the $Q$ factor being only in the thousands is mainly attributed to the fabrication errors in the hole diameter. By reducing the hole diameter error to a reasonable value, such as 4\%, the $Q$ factor can be significantly increased. Though challenging, such improvements in fabrication are still achievable.

Furthermore, we define a quantity, $A_\mathrm{eff}/A_\mathrm{uc}$, to quantify the enhancement tolerance to the QD position.
Here, $A_\mathrm{eff}$ represents the effective area within a unit cell where the LDOS exceeds half the maximum LDOS of the L20 PhC defect cavity. $A_\mathrm{uc}$ denotes the area of a single unit cell. As shown in Fig.~\ref{fig:S5}, the LDOS in moir\'e structures is remarkably higher than those in traditional cavities. Also, we have confirmed the average LDOS $\rho_\mathrm{eff}$ in the effective area $A_\mathrm{eff}$ for various cavities. Our analysis demonstrates that the moir\'e cavity exhibits a significantly enhanced average LDOS.

\section{Purcell enhancement of QD B}

To further validate the moir\'e cavity-enhanced QD fluorescence, we provide supplementary raw data from additional QD, labeled QD B, which is located in a moir\'e cavity adjacent to the one discussed in the main text, within the same chip. Figure~\ref{fig:S8}(a) shows spectra including the corresponding moir\'e cavity mode with the fluorescence emission from QD B. The dominant peak observed in the QD emission spectrum is attributed to QD B. Utilizing above-barrier excitation, we measure a fluorescence lifetime of 141$\pm$3~ps (Fig.~\ref{fig:S8}(b)), corresponding to a Purcell enhancement factor of approximately 8. Given the moir\'e cavity Q factor of 2191,  precisely tuning QD B in resonance with the cavity mode would predict a Purcell factor around 13.8, as shown as the peak value in Fig.~\ref{fig:S8}(c). We note that above-barrier excitation leads to long carrier relaxation time, typically hundreds of picoseconds~\cite{Berstermann2007,Reithmaier2014,Yan2023,Zibik2009}, from higher-energy states to the lowest exciton state, obscuring the true Purcell factor~\cite{Liu2018}. We anticipate that a reduced lifetime, hence a higher Purcell factor, can be measured if phonon-assisted or resonant excitation is employed, as demonstrated in the main text with QD A.

\newpage
\cleardoublepage

\begin{figure} 
\includegraphics[width=1\linewidth]{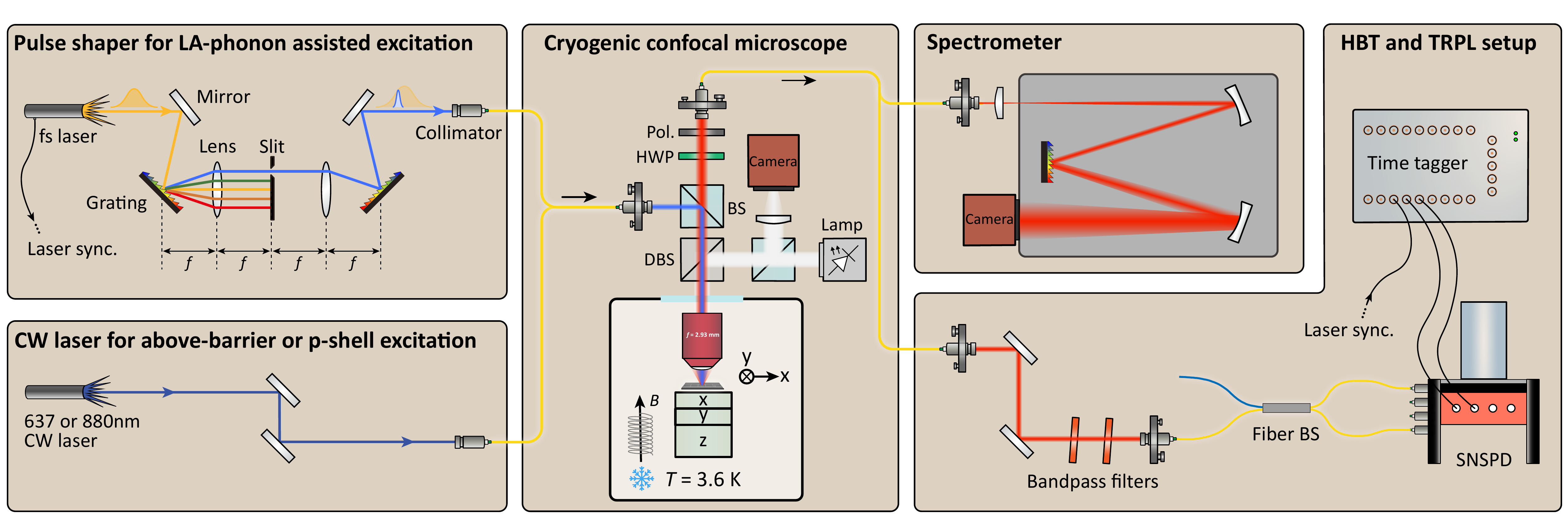}
    \caption{
    {\bf Schematic of the setup for optical measurements.} Left panels: laser excitation part including 4$f$ pulse shaping setup for LA-phonon-assisted excitation and CW lasers for above-barrier or p-shell excitation. Central panel: a home-built confocal microscope with the sample loaded in a closed-cycle cryostat (T = 3.6 K). Right panels: single-photon characterization part including spectrometer, HBT interferometer, and TRPL setup. Laser sync.: laser synchronization signal. Pol.: polarizer. HWP: half-wave plate. BS: beam splitter. DBS: dichroic beam splitter.
      } 
    \label{fig:Setup}
\end{figure}

\begin{figure} 
\centering
\includegraphics[width=0.95\linewidth]
{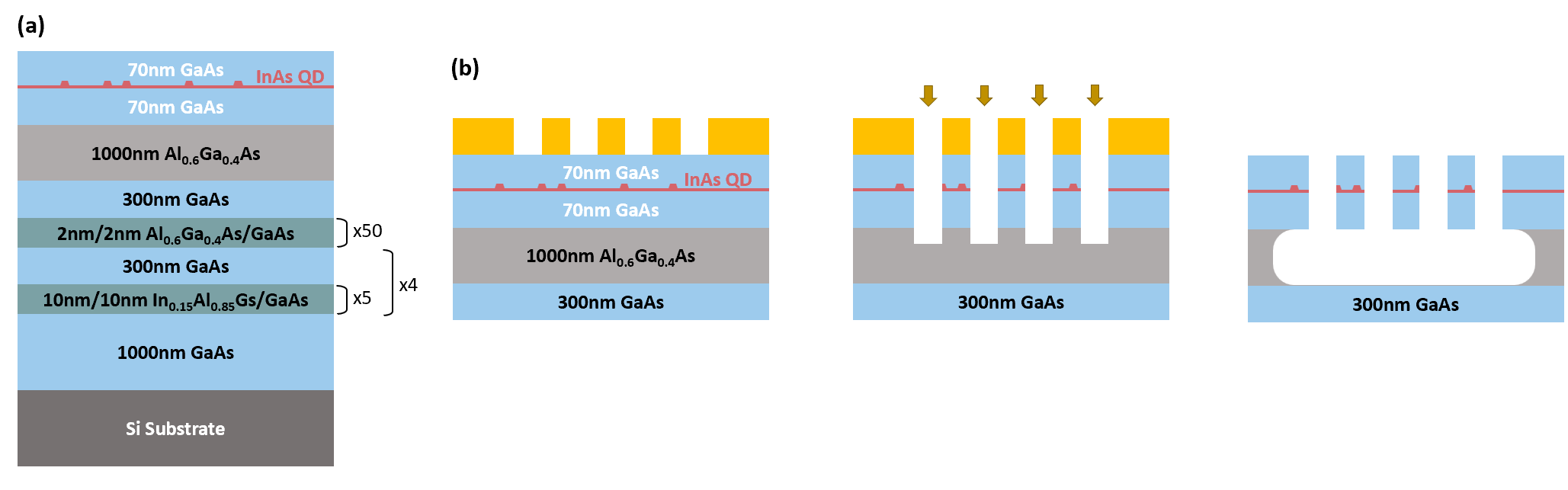}
    \caption{
    {\bf Wafer structure and sample fabrication process.} (a) Wafer structure of the InGaAs quantum dot sample. (b) Process flow for patterning and etching to achieve a suspended GaAs slab.
      } 
    \label{fig:S2}
\end{figure}

\begin{figure}
\centering
\includegraphics[width=0.95\linewidth]{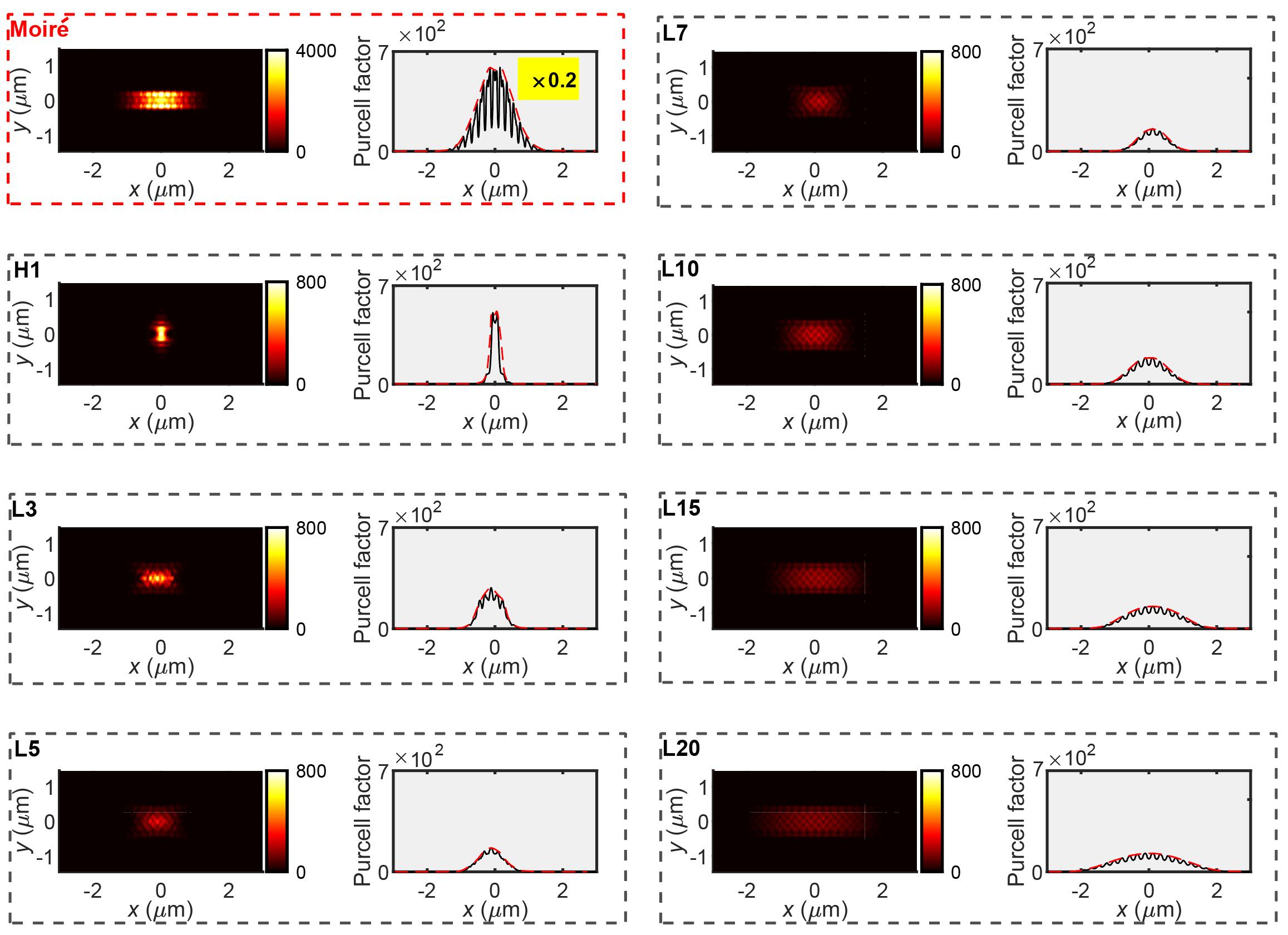}
    \caption{
    {\bf Distribution of the Purcell factor for moir\'e lattice, comparing to conventional defect PhC cavities.} The left column presents the distribution in the x-y plane. The right column is the corresponding result after averaging over the range y = -400nm to 400nm, with the red dashed line representing the envelope of its distribution. For visualization purposes, the averaged Purcell factor for the moir\'e structure is scaled down by a factor of 5. 
      } 
    \label{fig:S3}
\end{figure}

\begin{figure}   
\centering
\includegraphics[width=0.95\linewidth]{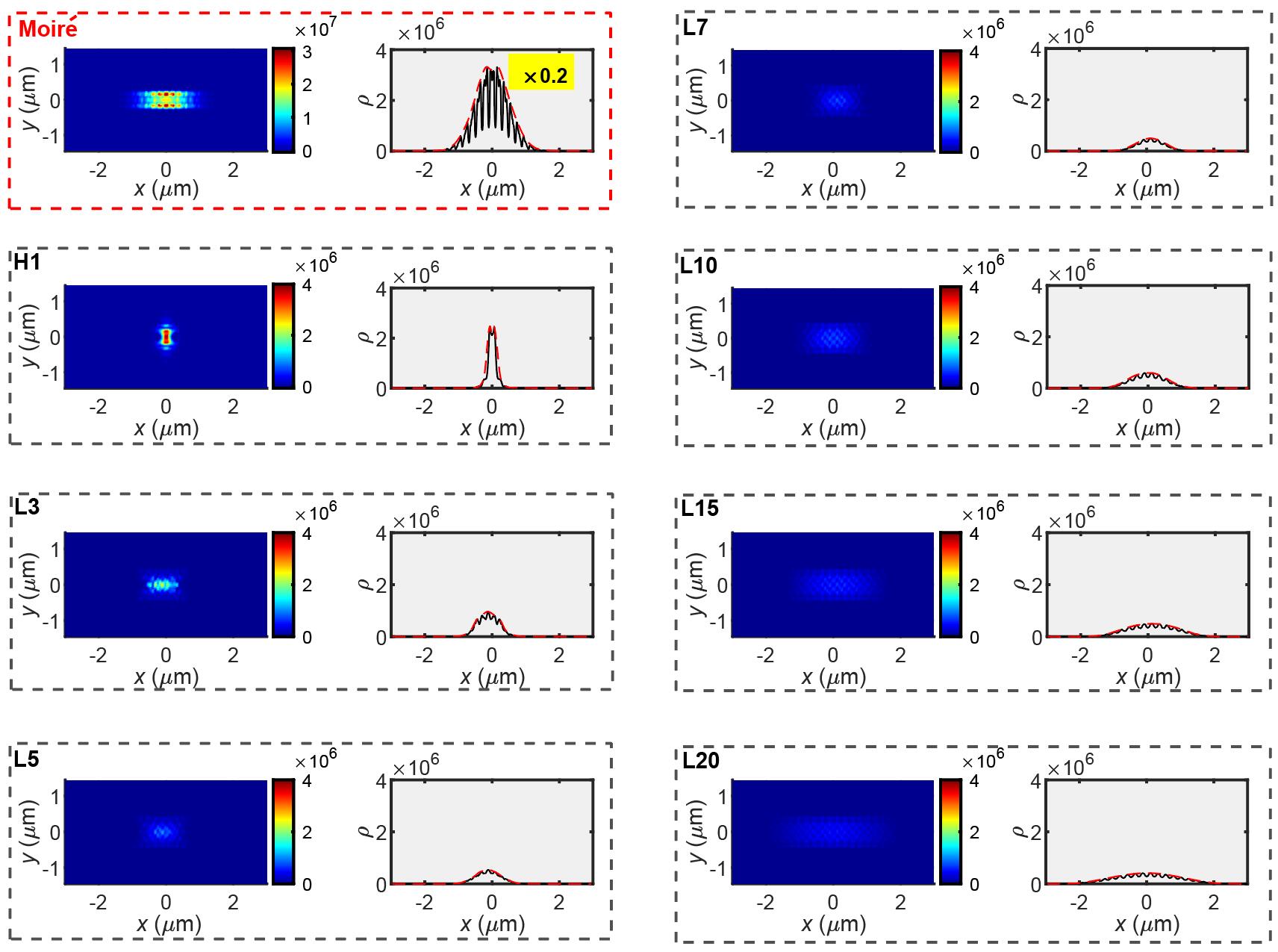}
    \caption{
    {\bf Distribution of the LDOS for moir\'e lattice, comparing to conventional defect PhC cavities.} The left column presents the distribution in the x-y plane. The right column is the corresponding result after averaging over the range y = -400nm to 400nm, with the red dashed line representing the envelope of its distribution. For visualization purposes, the averaged LDOS values for the moir\'e structure are scaled down by a factor of 5.
      } 
    \label{fig:S4}
\end{figure}

\begin{figure}
\centering
\includegraphics[width=.75\textwidth]{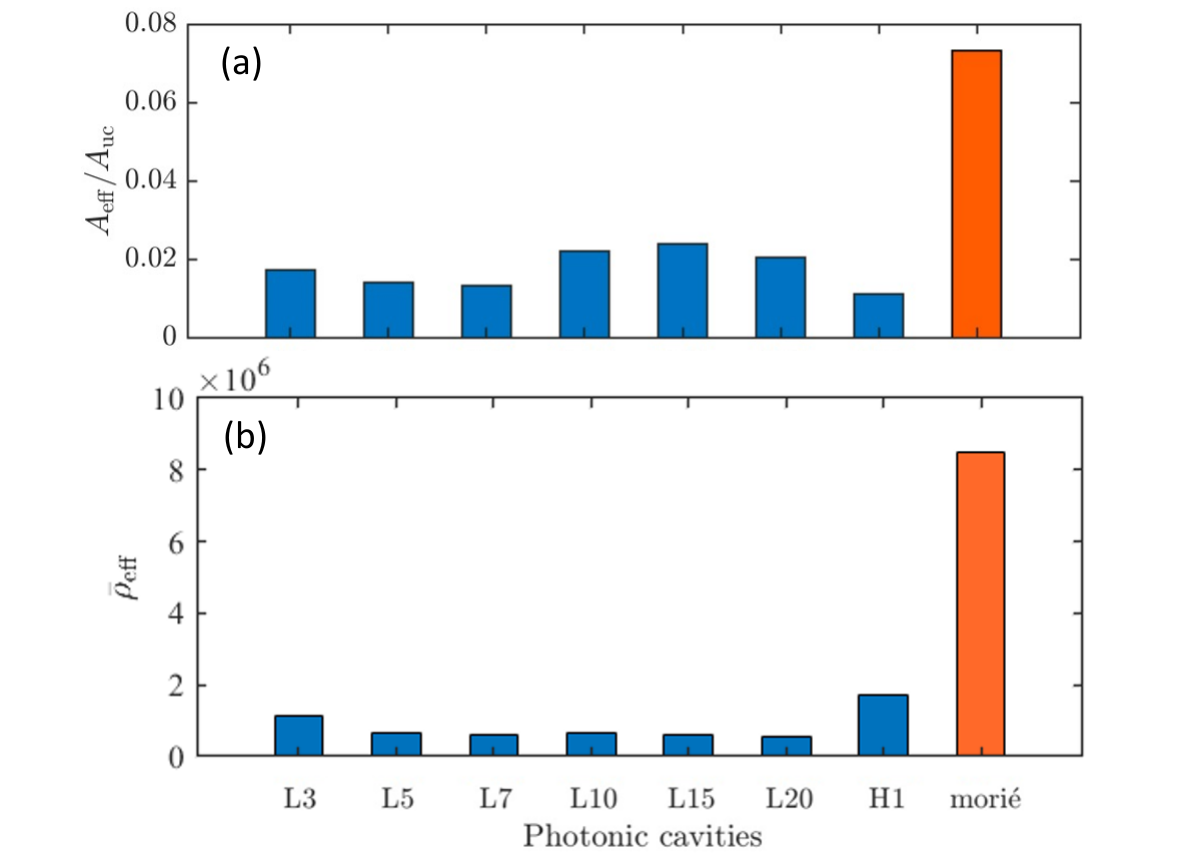}
\caption{
{\bf Numerical histogram results for main properties of various photonic cavities.} (a) Effective area for Aeff over unit cell area. (b) Average LDOS  on the effective area.
      }
      \label{fig:S5}
\end{figure}

\begin{figure}
\includegraphics[width=0.95\linewidth]{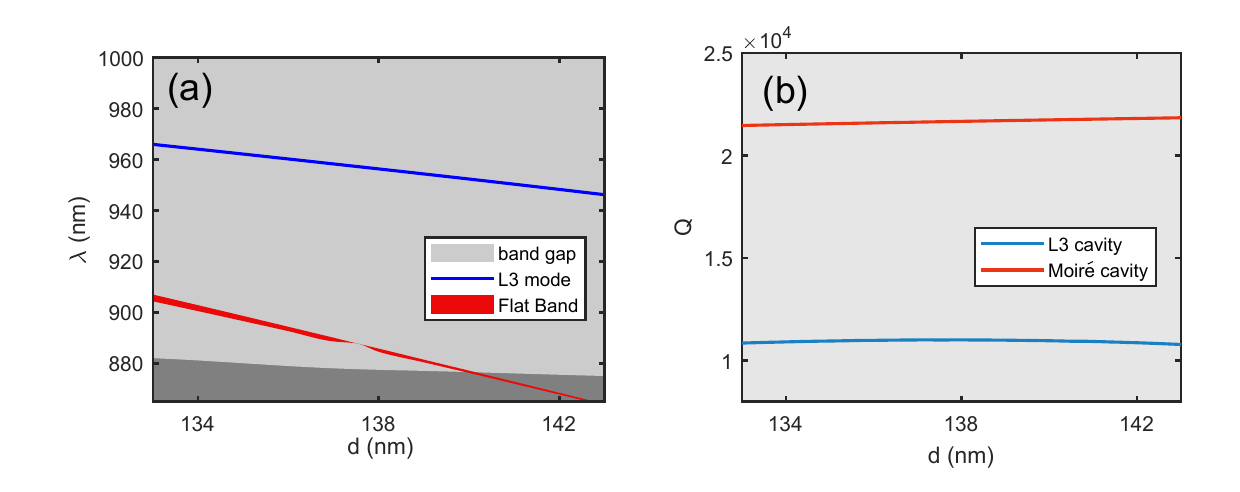}
    \caption{
    {\bf Numerical results for varying hole diameter.} (a) Frequency shift of flatband and L3 mode as a function of hole diameter. The red band denotes the flatband mode and its thickness represents the bandwidth. The dark grey region represents the bulk modes while the light grey region stands for the bandgap. (b) Q factor versus hole diameter for the moir\'e and L3 cavities. The Q factor of both cavities shows minimal variation with changes in hole diameter.
      } 
    \label{fig:S6}
\end{figure}

\begin{figure}[h]
\centering
\includegraphics[width=0.95\linewidth]{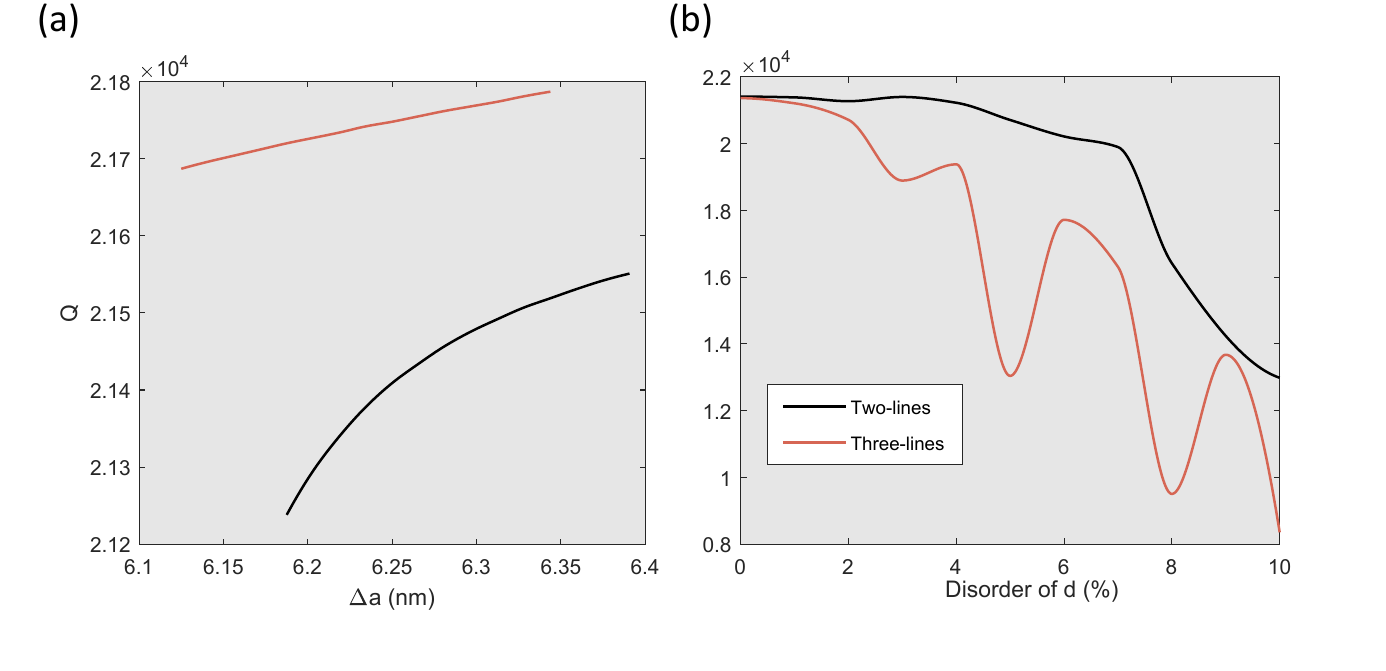}
\caption{
{\bf Robustness test for two designs of moir\'e structures.} (a) The Q factors vs. the lattice constant difference $\Delta a$ (b) The Q factors vs. Disorder of hole diameter $d$.
}
\label{fig:S7}
\end{figure}

\begin{figure} 
    \includegraphics[width=1\textwidth]{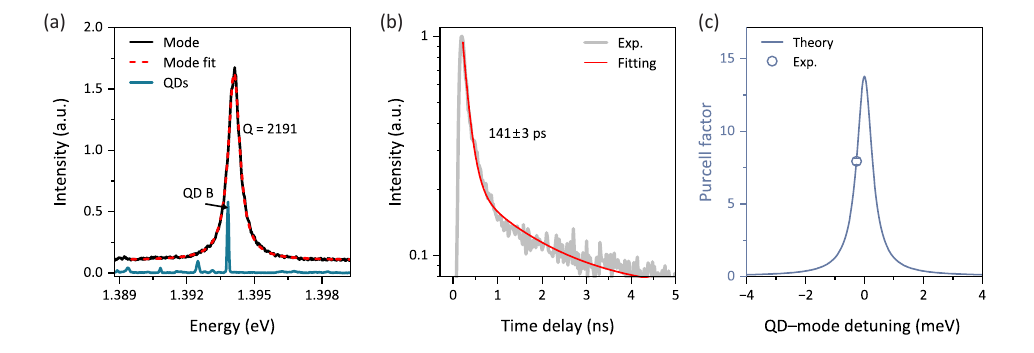}\\
    \caption{
    {
    \bf Purcell enhancement of QD B.
    }
    (a) Spectra of QD B and another moir\'e cavity mode. (b) Time-resolved PL of the QD B (c), The Purcell enhancement factor of the QD B as a function of detuning.
    }
    \label{fig:S8}
\end{figure}

\end{document}